\documentclass[12pt,a4paper,english]{article}
\usepackage{jheppub}

\usepackage[T1]{fontenc}
\usepackage[latin9]{inputenc}
\usepackage{amsmath}
\usepackage{graphicx}
\usepackage{amssymb}
\usepackage{esint}

\makeatletter

\providecommand{\tabularnewline}{\\}

\usepackage{graphics}

\usepackage{epsfig}\@ifundefined{definecolor}{\@ifundefined{definecolor}{\@ifundefined{definecolor}{\@ifundefined{definecolor}
 {\usepackage{color}}{}
}{}}{}}{}

\def\be{\begin{equation}}\def\ee{\end{equation}}\newcommand{\bea}{\begin{eqnarray}}\newcommand{\eea}{\end{eqnarray}}\def\bd{\begin{displaymath}}\def\ed{\end{displaymath}}\def\ba{\begin{eqnarray}}\def\ea{\end{eqnarray}}\definecolor{red}{rgb}{1,0,0}

\makeatother

\usepackage{babel}

\makeatother

\usepackage{babel}

\makeatother

\usepackage{babel}

\makeatother

\usepackage{babel}

\title{Can holography reproduce the QCD Wilson line?}
\author{Uri Kol}
\author{and Jacob Sonnenschein}

\affiliation{School of Physics and Astronomy,\\
The Raymond and Beverly Sackler Faculty of Exact Sciences,\\
Tel Aviv University, Ramat Aviv 69978, Israel}

\emailAdd{urikol@post.tau.ac.il}
\emailAdd{cobi@post.tau.ac.il}

\abstract{ Recently a remarkable agreement was found between lattice simulations
of long Wilson lines and behavior of the Nambu Goto string in flat
space-time. However, the latter fails to fit the short distance behavior
since it admits a tachyonic mode for a string shorter than a critical
length. In this paper we examine the question of whether a classical
holographic Wilson line can reproduce the lattice results for Wilson
lines of any length. We determine the condition on the the gravitational
background to admit a Coulombic potential at short distances. We analyze
the system using three different renormalization schemes. We perform
an explicit best fit comparison of the lattice results with the holographic
models based on near extremal $D_{3}$ and $D_{4}$ branes, non-critical
near extremal $AdS_{6}$ model and the Klebanov Strassler model. We
find that all the holographic models examined admit after renormalization
a constant term in the potential. We argue that the curves of the
lattice simulation also have such a constant term and we discuss its
physical interpretation.}

\begin{document}
\maketitle

\eject

\setcounter{page}{1}

\section{Introduction}

Recently lattice simulations of Wilson lines were found to admit a
remarkable agreement with the relation between the energy and the
length of the string as follows from the Nambu Goto (N.G.) formulation.
The simulations measured this relation in two \cite{Athenodorou:2007du}
and three \cite{Athenodorou:2010cs} space dimensions for both closed
strings wounded along a circle as well as for the open string case
\cite{Brandt:2010bw}. The dependence of the energy on the distance
between the endpoints of an open bosonic quantum string in $d$ dimensional
flat space-time is given by \cite{Arvis:1983fp} \be\label{NG}E=\sqrt{(\sigma \ell)^2+(n-\frac{d-2}{24})8\pi\sigma}\ee
where $\sigma$ is the string tension, $\ell$ is the separation distance
and $n$ is the excitation level. It was found out that for the ground
state the agreement with the full square-root formula is better than
the agreement with any approximation of this formula for long strings
($\sigma\ell^{2}>>1$), like $E\sim\sigma\ell+\frac{(d-2)}{24}\frac{4\pi}{\ell}+\frac{(d-2)}{24}\frac{8\pi^{2}}{\sigma\ell^{3}}+...\ $.
What is in particular remarkable is the fact that the agreement holds
also for short distances, up to $\ell\sqrt{\sigma}\sim1.6$.

Note however that the N.G. behavior (\ref{NG}) cannot serve as an
adequate description for distances shorter than the critical length
\be\label{NG.critical.length}\ell_{cr}= \frac{\pi(d-2)}{3} \frac{1}{\sqrt{\sigma}}\ee
since there the energy develops an imaginary part. It is thus clear
that to accommodate the short range behavior of the potential one
needs to go beyond the stringy picture of above. A potential solution
to this problem is to replace the string that resides in four dimensional
flat space-time with a holographic string in non-flat higher dimensional
space-time. An argument in favor of this scenario is the fact that
the behavior of the Wilson line in an $AdS_{5}$ space admits a $\frac{1}{\ell}$
which is similar to the behavior of the quark anti-quark potential
in short distances, namely $E\sim\frac{\alpha(\ell)}{\ell}$ where
$\alpha(\ell)$ is the running coupling. This naturally calls for a
holographic description in terms of a background that admits a confining
behavior \cite{key-1} for long strings and which is asymptotically
$AdS$.

Precise lattice simulations of the quark anti-quark potential especially
in the short distance range may serve as a test of the idea of weakly
curved holography. Whereas for long strings the contributions of massive
stringy modes are negligible, for short strings the impact of the
massive modes becomes more important. In holographic models massive
modes arise from fluctuations of the string along the radial direction,
compact directions and along fermionic directions. Thus high precision
profile for short distances will enable one to rule out or confirm
the idea of describing the Wilson line in terms of a string in weakly
curved higher dimensional space-time.

The goal of this paper is to search for a holographic background that
admits a Wilson line which is as close as possible to the potential
deduced from lattice calculations and which at the same time is in
accordance with other requirements. In \cite{key-1} the conditions
for a background to admit an area law behavior for long Wilson lines
were stated. For short Wilson lines the holographic model is required
to mimic the QCD perturbative behavior of the potential. To determine
this condition is a complicated task. Here instead, in order to have
a guide in the search of the appropriate background, we determine
the conditions for a background for which the short distance potential
behaves as $\frac{1}{\ell}$.

In the UV the lattice simulates QCD in its perturbative regime. The
holographic correspondence maps this regime to a highly curved gravitational
background. Thus obviously the comparison we make to the holographic
stringy result is on a solid ground only for large and maybe intermediate
distances. One may wonder whether any reliable information can be
extracted from the stringy picture in this case. Obviously the background
will be determined in this case by an action that includes higher
order curvature terms and moreover the stringy calculation will include
higher $\alpha'$ corrections. However, it is quite plausible that
the maximally symmetric $AdS$ background will remain a solution thought
the radius will differ from the classical value. For such a case,
even though the N.G. is not an adequate framework of calculation,
due to scale invariance the energy should scale as a $\frac{1}{\ell}$.

It is well known since the seminal paper \cite{Maldacena:1998im}
that the classical string action that corresponds to a Wilson line
diverges, and hence one has to invoke some renormalization procedure.
We discuss three different possible renormalization schemes: the holographic
renormalization of \cite{key-2}, the Legender transform approach
\cite{key-3} and the mass subtraction \cite{key-5}. It is shown
that for backgrounds without an horizon%
\footnote{ Horizon in general relativity associates with the time component
of the metric. Here and in the rest of the paper horizon will relate
to a space coordinate rather than to the time.%
} the three schemes yield the same result but for cases that have horizon
the last prescription differs from the two previous ones. From the
large arsenal of holographic models we have analyzed the models of
near extremal $D_{3}$ and $D_{4}$ branes \cite{Witten:1998zw},
non-critical near extremal $AdS_{6}$ \cite{key-6,Kuperstein:2004yf}
and the \textit{Klebanov Strassler} (\textit{KS}) background \cite{key-13} \footnote{Some similar aspects of the Wilson line in the KS background have already been discussed in \cite{Mia:2009wj}.}.

An outcome of this analysis is the fact that the quark anti-quark
potential admits on top of the linear and $\frac{1}{\ell}$ terms
also a constant term. This constant term appears in all the holographic
backgrounds we have analyzed and in all the renormalization schemes.
It also follows from the lattice simulations, where it is defined
as the difference between the constants added to find the best fit
with the simulation data in the large $\ell$ range and in the short
range. However, it seems that the constant term in the latter range
is not well-defined in field theory as the theoretical perturbative
expansion is not guaranteed to converge. Being aware of this point
we suggest different ways to discuss the constant term in spite of
this problem. The fact that the constants in the two different ranges
are different is what renders this constant to be a physical quantity.
A possible physical interpretation of this constant is in terms of
masses attached to the string endpoints.
A similar constant term has also been treated in the work of \cite{Hidaka:2009xh} on the zero point energy of renormalized Wilson loops, where the discussed action is a flat spacetime N.G. plus higher terms in covariant derivatives on the worldsheet.

Quantitative comparison between the lattice simulations and the holographic
results shows that the \textit{Klebanov Strassler} background exhibits
the smallest $\chi$ square. However, the non-critical $AdS_{6}$
model also admits a reasonable fit to the simulation results and it
has the advantage of being asymptotically $AdS$, while the $KS$
is only approximately asymptotically $AdS$ and hence doesn't protected
from quantum corrections in the $UV$. Both models are also invariant
under four dimensional Poincare symmetry, which is an essential requirement.

The paper is organized as follows. After this introduction, in section
\eqref{sec:The holographic Wilson line} we review the notion of Wilson
loops in holographic models. We then discuss the conditions on the
background to satisfy a coulombic potential at small distances. The
issue of the holographic renormalization is discussed in section \eqref{sub:Renormalization}.
We analyze separately the various renormalization schemes of \cite{key-2},
\cite{key-3} and \cite{key-5}. A brief review of certain holographic
models dual of Yang Mills like theories and discussion about their
Wilson line is presented in section \eqref{sec:the wilson line of the holographic
models}. In section \eqref{sec:Units-of-measurement} prior to the
performing of the comparison between the outcome of holography and
the results of lattice simulation, we setup the units appropriate
for the comparison. The latter is made in section \eqref{sec:Comparison-with-lattice}.
In section \eqref{sec:The-constant-terms-section} we discuss the
physical constant that follows from the simulations and the holographic
models. We end the paper in section \eqref{sec:Concluding-remarks}
with a brief summary and discussion of the findings of this paper
and certain open questions. In particular we discuss the issue of
the impact of quantum fluctuations on the holographic Wilson line.

\section{The holographic Wilson line\label{sec:The holographic Wilson line}}

We will investigate several holographic models, all of them are of
the form of some high dimensional space with a boundary that consists
some of the coordinates (from which only four should eventually be
infinite), a radial coordiante that takes us from the bulk of the
space to the boundary and can be viewed as energy scale and possibly
some other coordinates transverse to the boundary's coordinates. Following
\cite{key-1} we will assume that the metric depends only on the radial
coordinate such that its general form is \begin{equation}
ds^{2}=-G_{tt}\left(u\right)dt^{2}+G_{uu}\left(u\right)du^{2}+G_{x_{||}x_{||}}\left(u\right)dx_{||}^{2}+G_{x_{T}x_{T}}\left(u\right)dx_{T}^{2}\end{equation}
 where $t$ is the time direction, $u$ is the radial coordinate,
$x_{||}$ are the coordinates on the boundary and $x_{T}$ are the
transverse coordiantes. We adopt the notation in which the radial
coordinate is positive defined and the boundary is located at $u=\infty$.
In addition, an {}``horizon'' may exist at $u=u_{\Lambda}$, such
that spacetime is defined in the region $u_{\Lambda}<u<\infty$ instead
of $0<u<\infty$ in the case where no horizon is present. In order
to make our discussion most general, we will consider the case with
an existing horizon. Later on, when we imply the general analysis
to specific models, we would just set $u_{\Lambda}$ to zero in cases
with no horizon.

The construction we will examine is of an open string living in the
bulk of the space with its both ends tied to the boundary. From the
viewpoint of the field theory living on the boundary the endpoints
of the string are the $q\bar{q}$ pair, so the energy of the string
is related to the energy of the pair \cite{key-1}. In order to calculate
the energy of this configuration in the classical level we will use
the notations and results of \cite{key-1}. First, we define\begin{equation}
\begin{aligned}f^{2}\left(u\right) & \equiv G_{tt}\left(u\right)G_{x_{||}x_{||}}\left(u\right)\\
g^{2}\left(u\right) & \equiv G_{tt}\left(u\right)G_{uu}\left(u\right)\end{aligned}
\end{equation}
 Upon choosing the worldsheet coordinates $\sigma=x$ ($x$ is a coordinate
on the boundary pointing in the direction from one endpoint of the
string to the other one) and $\tau=t$ and assuming translation invariance
along $t$, the Nambu-Goto action describing the string takes the
form\begin{equation}
\begin{aligned}S_{NG} & =\int d\sigma d\tau\sqrt{det\left[\partial_{\alpha}X^{\mu}\partial_{\beta}X^{\nu}G_{\mu\nu}\right]}\\
 & =\int dxdt\sqrt{f^{2}\left(u\left(x\right)\right)+g^{2}\left(u\left(x\right)\right)\left(\partial_{x}u\right)^{2}}\\
 & =T\int dx\mathcal{L}\end{aligned}
\end{equation}
 Then the conjugate momentum and the Hamiltonian are\begin{equation}
p=\frac{\delta\mathcal{L}}{\delta u'}=\frac{g^{2}\left(u\right)u'}{\sqrt{f^{2}\left(u\right)+g^{2}\left(u\right)u'^{2}}}\label{eq:momentum}\end{equation}
 \begin{equation}
\mathcal{H}=p\cdot u'-\mathcal{L}=-\frac{f^{2}\left(u\right)}{\mathcal{L}}\end{equation}
 As the Hamiltonian does not depend explicitly on $x$, its value
is a constant of motion. We shall deal with the case in which $u\left(x\right)$
is an even function, and therefore there is a minimal value $u_{0}=u\left(0\right)$
for which $u'\left(0\right)=0.$ At that point we see from \eqref{eq:momentum}
that $p=0$. The constant of motion is therefore \begin{equation}
\mathcal{H}=-f\left(u_{0}\right)\end{equation}
 from which we can extract the differential equation of the geodesic
line \begin{equation}
\frac{du}{dx}=\pm\frac{f\left(u\right)}{g\left(u\right)}\frac{\sqrt{f^{2}\left(u\right)-f^{2}\left(u_{0}\right)}}{f\left(u_{0}\right)}\label{eq:geodesic}\end{equation}
 and re-express the on-shell lagrangian (i.e. the lagrangian on the
equation of motion) as a function of $f\left(u\right)$ only \begin{equation}
\mathcal{L}=\frac{f^{2}\left(u\right)}{f\left(u_{0}\right)}\end{equation}
 Then the distance between the string's endpoints (or the distance
between the {}``quarks'') would be\begin{equation}
\ell\left(u_{0}\right)=\int dx=\int du\left(\frac{du}{dx}\right)^{-1}=2\frac{1}{f\left(u_{0}\right)}\int_{u_{0}}^{u_{B}}du\frac{f^{2}\left(u_{0}\right)}{f^{2}\left(u\right)}\frac{g\left(u\right)}{\sqrt{1-\left(\frac{f\left(u_{0}\right)}{f\left(u\right)}\right)^{2}}}\label{eq:length}\end{equation}
 where $u_{0}$ is the minimal value in the radial direction to which
the string reaches and $u_{b}$ is the value of $u$ on the boundary.
The bare energy of the string is given by\begin{equation}
\begin{aligned}E_{bare}\left(u_{0}\right) & =\int dx\mathcal{L}=\int du\left(\frac{du}{dx}\right)^{-1}\mathcal{L}=2\int_{u_{0}}^{u_{b}}du\frac{g\left(u\right)}{\sqrt{1-\left(\frac{f\left(u_{0}\right)}{f\left(u\right)}\right)^{2}}}\\
 & =f\left(u_{0}\right)\cdot\ell\left(u_{0}\right)+2\int_{u_{0}}^{u_{b}}du\, g\left(u\right)\sqrt{1-\frac{f^{2}\left(u_{0}\right)}{f^{2}\left(u\right)}}\end{aligned}
\label{eq:bareEnergy}\end{equation}
 Generically, the bare energy diverges and hence a renormalization
procedure is needed. Possible renormalization schemes are discussed
in section \eqref{sub:Renormalization}. Here we follow \cite{key-1}
and use the mass subtraction scheme in which the bare masses of the
quarks are subtracted from the bare energy. The bare quark mass is
viewed as a straight string with a constant value of $x$, stretching
from $u=0$ (or $u=u_{\Lambda}$ if there exists an horizon at $u_{\Lambda}$)
to $u=u_{b}$, such that it is given by \begin{equation}
m_{q}=\int_{u_{\Lambda}}^{u_{b}}du\, g\left(u\right)\end{equation}
 Then the renormalized energy would be given by\begin{equation}
E\left(u_{0}\right)=f\left(u_{0}\right)\cdot\ell\left(u_{0}\right)-2\mathcal{K}\left(u_{0}\right)\label{eq:Energy}\end{equation}
 where $\mathcal{K}\left(u_{0}\right)$ is\begin{equation}
\mathcal{K}\left(u_{0}\right)=\int_{u_{0}}^{u_{b}}du\, g\left(u\right)\left(1-\sqrt{1-\frac{f^{2}\left(u_{0}\right)}{f^{2}\left(u\right)}}\right)+\int_{u_{\Lambda}}^{u_{0}}du\, g\left(u\right)\label{eq:Kappa}\end{equation}
 In section \eqref{sub:Renormalization} we make a discussion about
the differences between different renormalization schemes. We emphasize
that this result for the energy is only at the classical level and
do not include quantum corrections.

In order to reproduce the QCD heavy quarks potential we first have
to demand the holographic models to reproduce the asymptotic forms
of the potential. This leads to several restrictions on the forms
of the $f$ and $g$ functions. The condition for confining behavior
at large distances was derived in \cite{key-1}:
\begin{enumerate}
\item $f$ has a minimum at $u_{min}$ and $f\left(u_{min}\right)\neq0$
or
\item $g$ diverges at $u_{div}$ and $f\left(u_{div}\right)\neq0$
\end{enumerate}
Then the string tension is given by $f\left(u_{min}\right)$ or $f\left(u_{div}\right)$,
correspondingly. The second asymptotic is perturbative QCD at small
distances. In section \eqref{sec:Asymptotically Coulombic} we derive
the conditions on the background to reproduce the leading perturbative
behavior of QCD, which is Coulomb-like (throughout this paper we refer
to a $\frac{1}{\ell}$ behavior as Coulombic).

The physical picture arising from this construction is as follows.
The confining limit is approached as $u_{0}\rightarrow u_{\Lambda}$,
then most of the string lies on the horizon%
\footnote{For simplicity we depict the picture only for the case where horizon
is present, but a similar picture arise also in cases with no horizon.
For more extensive discussion see \cite{key-20}.%
} and implies a string-like interaction between the {}``quarks''
(the endpoints of the string) with a string tension \begin{equation}
\sigma=f\left(u_{\Lambda}\right)\label{eq:string tension}\end{equation}
 and a linear potential\begin{equation}
E=\sigma\cdot\ell-2\kappa\end{equation}
 where $\kappa=\mathcal{K}\left(u_{\Lambda}\right)$ is a finite constant.
On the other hand, the conformal limit is approached as $u_{0}/u_{\Lambda}\rightarrow\infty$,
then the whole string is far away from the horizon and it is ruled
by the geometry near the boundary. In other words, this limit correspondes
in some sense to taking $u_{\Lambda}$ to zero, such that the scale
parameter disappears and the conformal theory is reconstructed.

\begin{figure}
\includegraphics{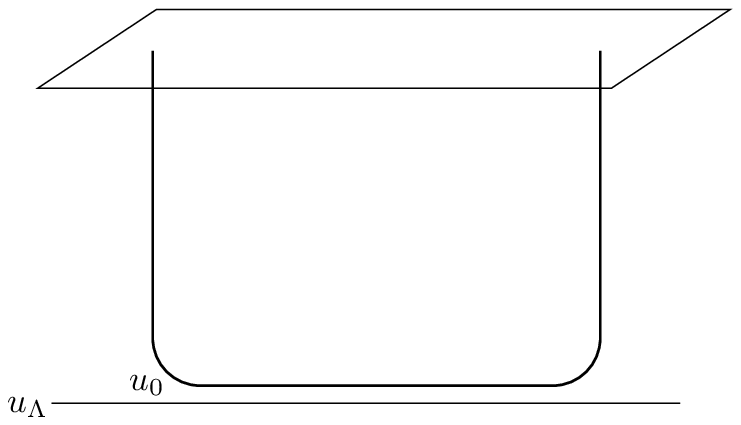}\includegraphics{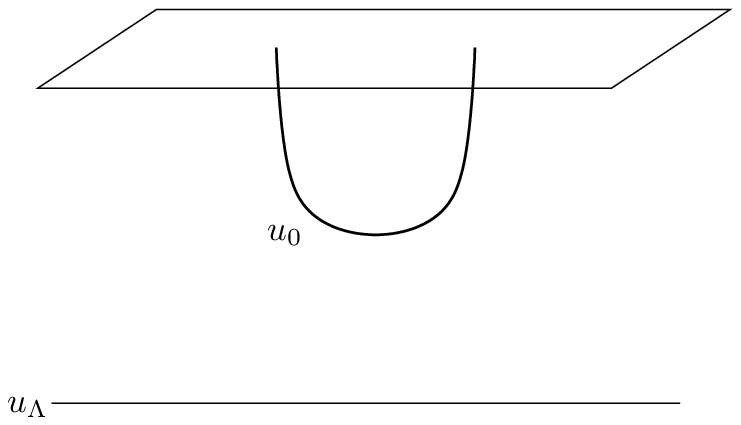}\caption{The physical picture arising from the present construction. Left:
The confining limit is approached as the string is close to the horizon,
then most of the string lies on the horizon and implies a string-like
interaction between the quarks (the endpoints of the string). Right:
The conformal limit is reconstructed when the string is far away from
the horizon such that it is ruled by the geometry near the boundary.}

\end{figure}

\section{The conditions on the background to satisfy a coulombic potential
at small distances\label{sec:Asymptotically Coulombic}}

In order to generate the QCD heavy quarks potential in an holographic
model it should first reconstruct the potential at asymptotics. The
conditions on the holographic background to reproduce the confining
asymptotic at large distances were derived in \cite{key-1}. As was
discussed in the introduction, the holographic dual of the perturbative
region is necessarily highly curved and hence unreliable. Nevertheless
we would like to derive the conditions on the background to reproduce
at small distances a potential of the form\begin{equation}
E\sim\frac{1}{\ell}\end{equation}
 The full derivation is detailed in appendix \eqref{sec:asymptotically coulombian appendix}.

The first step would be to expand the $f$ and $g$ functions around
the boundary, which we will take to be at infinity $u_{b}=\infty$,\begin{equation}
\begin{aligned}f\left(u\right) & =a_{k}u^{k}+\mathcal{O}\left(u^{k-1}\right)\\
g\left(u\right) & =g\left(\infty\right)+b_{j}u^{-j}+\mathcal{O}\left(u^{-j+1}\right)\\
k,j & >0\end{aligned}
\label{eq:ansatz.asy.Coulombic}\end{equation}
 We assume here that $f\left(u\right)$ diverges on the boundary as
a power law and that $g\left(u\right)$ tends to a finite value on
the boundary. Then we can calculate the length of the string as a
function of its minimal radial position, $u_{0}$, using \eqref{eq:length}.
Since we are interested in the small distance behavior, i.e. $u_{0}$
which is close to the boundary, the calculation will be done in the
region $u_{0}\gg u_{\Lambda}$, there we find\begin{equation}
\ell\left(u_{0}\right)=\left(\frac{2g\left(\infty\right)}{a_{k}}C_{2k,2k}\left(\infty\right)\right)\frac{1}{\left(u_{0}\right)^{k-1}}+\mathcal{O}\left(\frac{1}{\left(u_{0}\right)^{k+j-1}}\right)\label{eq:length2}\end{equation}
 where\begin{equation}
C_{2k,2k}\left(\infty\right)=\int_{1}^{\infty}dy\frac{1}{\left(y\right)^{2k}}\frac{1}{\sqrt{1-y^{-2k}}}\end{equation}
 is a convergent integral. Next we calculate $\mathcal{K}\left(u_{0}\right)$
in the same manner\begin{equation}
\mathcal{K}\left(u_{0}\right)=u_{0}g\left(\infty\right)\left[1+\frac{1}{2}C'_{2k,2k}\left(\infty\right)\right]+\mathcal{O}\left(u_{0}^{1-j}\right)\end{equation}
 (where $C'_{2k,2k}\left(\infty\right)$ is another convergent integral
defined in appendix \eqref{sec:asymptotically coulombian appendix})
and use \eqref{eq:length2} to find the inverse relation of $u_{0}$
as a function of $\ell$ \begin{equation}
u_{0}=\left(\frac{2g\left(\infty\right)}{a_{k}}C_{2k,2k}\left(\infty\right)\frac{1}{\ell}\right)^{\frac{1}{k-1}}\end{equation}
 in order to express $\mathcal{K}$ as a function of the length\begin{equation}
\mathcal{K}=\left[1+\frac{1}{2}C'_{2k,2k}\left(\infty\right)\right]\left[2C_{2k,2k}\left(\infty\right)\right]^{\frac{1}{k-1}}\left[g\left(\infty\right)\right]^{\frac{k}{k-1}}\left(\frac{1}{a_{k}\ell}\right)^{\frac{1}{k-1}}\end{equation}
 Then we are ready to substitute these results into the expression
for the energy of the string \eqref{eq:Energy} as a function of its
length to get\begin{equation}
\begin{aligned}E\left(\ell\right) & =f\left(u_{0}\left(\ell\right)\right)\cdot\ell-2\mathcal{K}\left(\ell\right)\\
 & =N_{k}\left[g\left(\infty\right)\right]^{\frac{k}{k-1}}\left(\frac{1}{a_{k}\ell}\right)^{\frac{1}{k-1}}\end{aligned}
\end{equation}
 where $N_{k}$ is a $k$-dependent dimensionless number\begin{equation}
N_{k}=\left[1+2C_{2k,2k}\left(\infty\right)+\frac{1}{2}C'_{2k,2k}\left(\infty\right)\right]\left[2C_{2k,2k}\left(\infty\right)\right]^{\frac{1}{k-1}}\end{equation}
 The bottom line is that the dependence of the energy on the length
in the region close to the boundary, i.e. small distances, is

\begin{equation}
E\sim\left(\frac{1}{\ell}\right)^{\frac{1}{k-1}}\end{equation}
 We conclude that in order for the potential to satisfy a coulombic
form near the boundary, one has to demand $k=2$. In other words,
the condition on the backround is

\begin{equation}
f\left(u\gg u_{\Lambda}\right)\sim u^{2}\label{eq: condition.coulombic.behavior}\end{equation}
 We emphasize that this analysis is valid as long as $f$ diverges
on the boundary as a power law and $g$ is finite there. An example
for a case that fulfills this ansatz is the famous $AdS_{5}\times S_{5}$
solution in which\begin{equation}
\begin{aligned}f\left(u\right) & =\frac{1}{2\pi\alpha'}\left(\frac{u}{R_{AdS}}\right)^{2}\\
g\left(u\right) & =\frac{1}{2\pi\alpha'}\end{aligned}
\end{equation}

\section{Renormalization\label{sub:Renormalization}}

Generically, the bare energy of the string in the holographic construction
\eqref{eq:bareEnergy} is divergent. From the bulk perspective it
is a result of the infiniteness of the boundary. The string's endpoints
are tied to the boundary and the energetically preferred state is
when the string extends to the bulk%
\footnote{There may be gravitational solutions in which this is not the case
and the energetically preferred state is when the whole string extends
to the boundary only, but then the Wilson loop is just the flat space
Wilson loop. We do not consider such a case as we are interested in
holographic Wilson loops.%
}. Since the boundary is radialy located at infinity, this would mean
that the string's length is infinite, and this is the origin of the
divergence. From the field theory living on the boundary perspective
the infinite bare mass of the quarks is the source for the divergence.
The most simple and naive approach would be to subtract this bare
mass \cite{Maldacena:1998im}, as one can think of the quark as a
straight string stretched from $u=0$ (or $u=u_{\Lambda}$ if there
exist an horizon at $u_{\Lambda}$) to $u=u_{b}$%
\footnote{In this section we will keep the notation in which the boundary is
located at $u=u_{b}$. Later on we will set $u_{b}=\infty$.%
} such that its mass is given by \begin{equation}
m_{q}=\int_{u_{\Lambda}}^{u_{b}}du\, g\left(u\right)\end{equation}
 That approach was taken in \eqref{sec:The holographic Wilson line}
following \cite{key-1}. However, there may be other regularization
schemes as well, and we want to discuss and compare them. In this
section we examine two other schemes: the holographic scheme \cite{key-2}
and the Legendre transform scheme \cite{key-3}. We calculate analytically
the subtracted part of the energy in each of the schemes, and then
discuss the results.

The calculation is done using the expression for the bare energy \eqref{eq:bareEnergy}
and assuming the following assumptions on the $f$ and $g$ functions:
\begin{enumerate}
\item $f$ is a monotonic increasing function
\item $g$ is a constant function or a monotonic decreasing function which
does not diverge on the boundary
\end{enumerate}
The first one is the same as the ansatz for the $f$ function in \cite{key-1},
while the second ansatz is more restrictive, but in fact is not needed
in our analysis of the Legendre transform scheme.

\subsection{Holographic renormaliztion}

The idea behind this scheme \cite{key-2} is to expand the bare energy
around the boundary, identify the infinite terms and subtract them
from the original expression of the bare energy. Essentialy we remove
infinities resulting from the infiniteness of the boundary. Therefore
we will not lose any information about the subtracted part if we will
consider the bare energy with $u_{0}$ which is far from the horizon.
Then $g\left(u\right)$ is approximately constant in all the range
of integration

\begin{equation}
E_{bare}\left(u_{0}\right)\cong2g\left(u_{b}\right)\int_{u_{0}}^{u_{b}}du\frac{1}{\sqrt{1-\left(\frac{f\left(u_{0}\right)}{f\left(u\right)}\right)^{2}}}\end{equation}
 Now let us divide the integration into two parts\begin{equation}
E_{bare}\left(u_{0}\right)=2g\left(u_{b}\right)\left[\int_{u_{m}}^{u_{b}}du\frac{1}{\sqrt{1-\left(\frac{f\left(u_{0}\right)}{f\left(u\right)}\right)^{2}}}+\int_{u_{0}}^{u_{m}}du\frac{1}{\sqrt{1-\left(\frac{f\left(u_{0}\right)}{f\left(u\right)}\right)^{2}}}\right]\end{equation}
 such that $f\left(u\right)\gg f\left(u_{0}\right)$ above $u_{m}$
so we can simplify this expression\begin{equation}
\begin{aligned}E_{bare}\left(u_{0}\right) & \cong2g\left(u_{b}\right)\left[\int_{u_{m}}^{u_{b}}du+\int_{u_{0}}^{u_{m}}du\frac{1}{\sqrt{1-\left(\frac{f\left(u_{0}\right)}{f\left(u\right)}\right)^{2}}}\right]\\
 & =2u_{b}g\left(u_{b}\right)-2g\left(u_{b}\right)u_{0}\left[\frac{u_{m}}{u_{0}}-\int_{1}^{\frac{u_{m}}{u_{0}}}dy\frac{1}{\sqrt{1-\left(\frac{f\left(u_{0}\right)}{f\left(u\right)}\right)^{2}}}\right]\end{aligned}
\end{equation}
 where in the last transition we have changed the integration variable
to $y=\frac{u}{u_{0}}.$ The second term does not contains boundary
infinities, because its only dependence on the boundary is via $g\left(u_{b}\right)$
which is finite in our ansatz. So even if the second term includes
divergent parts, this scheme will not take care of it. Writing the
bare energy in that form we isolate the infinity resulting from the
boundary such that according to the prescription of this scheme, the
subtracted part of the energy would be\begin{equation}
E_{sub}=2u_{b}g\left(u_{b}\right)\end{equation}

\subsection{The Legendre transform scheme}

The idea behind this scheme \cite{key-3} is to consider the legendre
transform of the energy as the renormalized energy\begin{equation}
E_{renormalized}=\int dx\left(\mathcal{L}-p\cdot u'\right)\end{equation}
 such that the subtracted part of the energy is\begin{equation}
E_{sub}=\int dx\left(p\cdot u'\right)\end{equation}
 Now, since $p$ does not depend explicitly on $x$ (see \eqref{eq:momentum})
we can write it as\begin{equation}
E_{sub}=\int dx\frac{\partial}{\partial x}\left(p\cdot u\right)\end{equation}
 which means that the subtracted part of the energy is a boundary
term, and therefore we can expect to have a similar result as in the
holographic scheme. Changing variables from $x$ to $u$ ($dx\frac{\partial}{\partial x}=du\frac{\partial}{\partial u}$)
and substitute the momentum \eqref{eq:momentum} on the geodesic line
\eqref{eq:geodesic}, we have

\begin{equation}
\begin{aligned}E_{sub} & =\int_{u_{0}}^{u_{b}}du\frac{\partial}{\partial u}\left[2ug\left(u\right)\sqrt{1-\left(\frac{f\left(u_{0}\right)}{f\left(u\right)}\right)^{2}}\right]\\
 & =2u_{b}g\left(u_{b}\right)\sqrt{1-\left(\frac{f\left(u_{0}\right)}{f\left(u_{b}\right)}\right)^{2}}\\
 & =2u_{b}g\left(u_{b}\right)\end{aligned}
\end{equation}
 which indeed result as the holographic scheme. Pay special attention
to the fact that we have not used here the ansatz for the $g$ function
stated in the beginning of this section and still got the same result
as for the holographic scheme.

\subsection{The mass subtraction scheme\label{sub:The-Mass-Subtraction}}

This scheme suggests to simply subtract the bare mass of two free
quarks, each of them is represented by a straight string stretched
from the boundary to the horizon. Therefore the subtracted energy
is given by \cite{key-1}

\begin{equation}
E_{sub}=2\int_{u_{\Lambda}}^{u_{b}}du\, g\left(u\right)\end{equation}
 Let us seperate the integration into two regions\begin{equation}
E_{sub}=2\int_{u_{m}}^{u_{b}}du\, g\left(u\right)+2\int_{u_{\Lambda}}^{u_{m}}du\, g\left(u\right)\end{equation}
 where above $u_{m}$, $g$ is approximately stabilized on its boundary
value\begin{equation}
\begin{aligned}E_{sub} & \cong2g\left(u_{b}\right)\int_{u_{m}}^{u_{b}}du+2\int_{u_{\Lambda}}^{u_{m}}du\, g\left(u\right)\\
 & =2u_{b}g\left(u_{b}\right)-u_{\Lambda}\left[2\frac{u_{m}}{u_{\Lambda}}g\left(u_{b}\right)-2\int_{1}^{\frac{u_{m}}{u_{\Lambda}}}dy\, g\left(y\right)\right]\end{aligned}
\end{equation}
 where in the last transition we have changed the integration variable
to $y=\frac{u}{u_{\Lambda}}.$ The first term is essentialy the same
one as of the two other schemes, and is a result of the infiniteness
of the boundary. The second term is a feature of the horizon and it
is finite since it does not contain boundary infinities (because we
isolated them) and its only dependence is on the horizon characteristics,
which cannot cause a divergence of the energy (because the curvature
is not singular). Its exact value can be calculated for a given model.
We conclude that the subtracted energy in this scheme takes the form\begin{equation}
E_{sub}=2u_{b}g\left(u_{b}\right)-A\cdot u_{\Lambda}\end{equation}
 where $A$ is a constant finite number that depends on the exact
form of the horizon\begin{equation}
A=lim_{\frac{u_{m}}{u_{\Lambda}}\rightarrow\infty}\left[2\frac{u_{m}}{u_{\Lambda}}g\left(u_{b}\right)-2\int_{1}^{\frac{u_{m}}{u_{\Lambda}}}dy\, g\left(y\right)\right]\end{equation}
 In the non critical $AdS_{n+1}$ solutions that will be discussed
later, for example, one finds\begin{equation}
A\left(n\right)=\frac{\Gamma\left[\frac{n-1}{n}\right]}{\sqrt{\pi}\Gamma\left[\frac{n-2}{2n}\right]}\end{equation}

\begin{figure}
\includegraphics{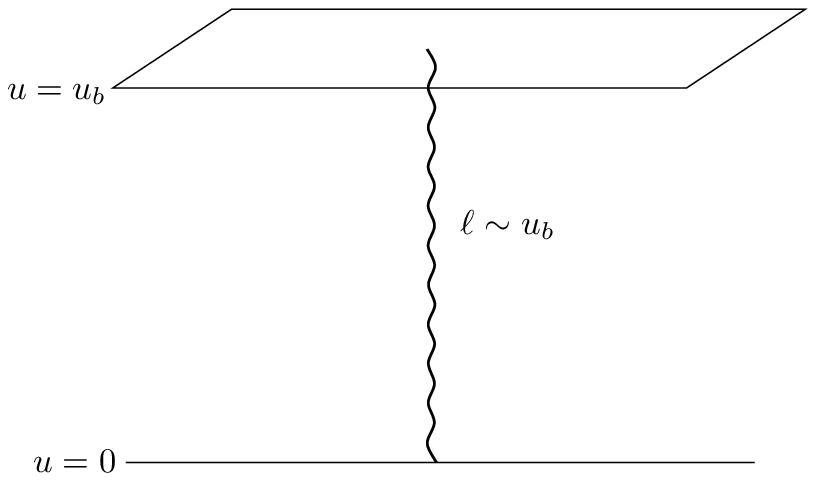}\includegraphics{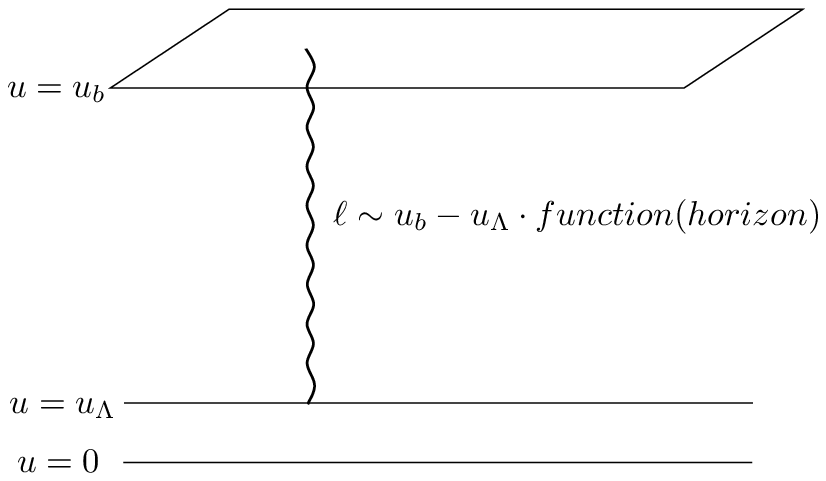}

\caption{\label{fig:reno.schemes}Left: In the case in which the renormalization
scheme ignores the horizon, one can think of this scheme as of subtracting
the length of a string streched from the boundary to the end of space
at $u=0$, which is proportional to the location of the horizon, $u_{b}$.
Right: In the mass subtraction scheme one subtracts the length of
a string streched from the boundary to the horizon, which is shorter
by a finite amount from the string of the first case, as the effect
of the horizon must be finite. On dimensional grounds, this finite
difference must be proportional to $u_{\Lambda}$ and it depends only
on the characteristics of the horizon ($function\left(horizon\right)$
is a finite dimensionless number).}

\end{figure}

\subsection{Conclusions}

We see that the mass subtraction scheme differs from the two other
schemes by a constant term in the expression for the subtracted energy.
The fact that different schemes differ only by a finite term is true
in field theory and here we have a geometrical realization for that
feature, as explained below.

The holographic and Legendre schemes are taking care only of the infintiness
of the boundary, and do not take into account the existence of the
horizon, while the mass subtraction scheme does take it into account.
Indeed, in cases with no horizon we set $u_{\Lambda}$ to zero (according
to the prescription in section \eqref{sec:The holographic Wilson line})
and the difference between the schemes disappears. As was mentioned
in the beginning of this section, the divergence is a result of the
infiniteness of the boundary. The difference between the schemes is
only the horizon, which cannot cause any divergences of the energy
since the curvature is not singular. Therefore we conclude that the
difference between the schemes must be realized as a finite affect.

An important implication of this conclusion is that the analysis of
the behavior of the Wilson loop at asymptotic regions (in \cite{key-1}
for the confining region and in section \eqref{sec:Asymptotically Coulombic}
for the coulombic region) is independent of the regularization scheme
(as far as those three schemes are considered).

Furthermore, out of the fact that the $UV$ region is eventually decoupled
from the $IR$ region as we take the limit $u_{b}\rightarrow\infty$,
one can restrict the form of the subtracted parts. Each of those regions
has its own scale parameter ($u_{b}$ and $u_{\Lambda}$, respectively)
and therefore properties of each of them would depend only on the
relevant scale. The divergent part of the energy is a feature of the
$UV$ region (since the divergency is a result \textit{only} of the
infiniteness of the boundary) such that on dimensional grounds it
would be proportional to the location of the boundary, $u_{b}$. On
the other hand, the finite difference between the schemes would be
proportional to $u_{\Lambda}$ since it is a result of the $IR$ properties
of the background. We describe this picture in figure \eqref{fig:reno.schemes}.

\section{The Wilson line of the holographic models\label{sec:the wilson line of the holographic models}}

So far we have made a general discussion about the Wilson line in
holographic backgrounds. As we aim to compare the QCD Wilson line
to the one of holographic theories, we have to consider specific models.
In this section we briefly review few representative holographic models,
which we latter compare to lattice simulations, and discuss their
Wilson line. Among the large class of holographic models we have choose
to consider the near extremal $D_{3}$ and $D_{4}$ branes, non critical
$AdS_{6}$ and the \textit{Klebanov Strassler} construction. The reason
for this choice is that these backgrounds admit a Wilson line which
is in principle close to the QCD Wilson line (at least on asymptotics),
as we will discuss below. In addition, they are all solutions of string
theory or of supergravity. Other examples have been analyzed in the
literature, among them also phenomenological models and backgrounds
which are not solutions of a particular action. In this paper we choose
not to deal with such cases since one can always engineer a background
which will admit some required features without being a solution of
an action. Instead, we want to ask what is the \textit{solution} that
admits a Wilson line which is closest to the QCD Wilson line. Nevertheless,
for a matter of completeness, we shortly review several other results;
The Wilson line of the phenomenological Holographic QCD \cite{key-21}
has been analyzed in \cite{key-22} where a good agreement to the
lattice results was found. In \cite{key-15}, the Wilson line of some
hard-wall scenario was calculated and it was found to admit a Cornell
potential form%
\footnote{We thank Andrew Zayakin for referring us to this paper.%
}.

\subsubsection*{Near extremal $D_{p}$ brane}

In the supergravity background associated to near extremal $D_{p}$
brane \cite{key-4} one finds\begin{equation}
\begin{aligned}f^{2}\left(u\right) & =\left(\frac{1}{2\pi\alpha'}\right)^{2}\left(\frac{u}{R_{AdS}}\right)^{7-p}\\
g^{2}\left(u\right) & =\left(\frac{1}{2\pi\alpha'}\right)^{2}\frac{1}{1-\left(\frac{u_{\Lambda}}{u}\right)^{7-p}}\end{aligned}
\end{equation}
 These backgrounds satisfy the condition for confinement for all values
of $p$. On the other hand, using the analysis of \eqref{sec:Asymptotically Coulombic},
we see that the potential in the small distances region takes the
form\[
E\sim\left(\frac{1}{\ell}\right)^{\frac{2}{5-p}}\]
 Therefore, only for $p=3$ the potential behaves as $\frac{1}{\ell}$
in the small distances region. In that case there are four coordinates
on the boundary, where one of them is compactified. For that reason,
on energy scales smaller than one over the compactification radius\begin{equation}
E\lesssim\frac{1}{R_{compactification}}\end{equation}
 the theory on the boundary is effectively three dimensional. The
conclusion is that the field theory on the boundary of the near extremal
$D_{3}$ brane is four dimensional at small lengths scale (or equivalently
high energies scale), where the potential is Coulombic, but it is
effectively three dimensional on large lengths scale (low energies),
where the potential is confining. In order to finish with an effectively
four dimensional boundary at large distances, one needs to consider
the near extremal $D_{4}$ brane, following the same argument. In
that case the boundary theory on small length scales would be five
dimensional and the potential there will take the form $\sim\frac{1}{\ell^{2}}$.

\subsubsection*{Non critical $AdS$ solutions}

The metric of a non-critical $AdS_{n+1}$ solution of type II string
theory \cite{key-6,Kuperstein:2004yf} with one of the spatial coordinates
compactified on a circle is given by \begin{equation}
\begin{aligned}ds^{2} & =\frac{1}{2\pi\alpha'}\left(\frac{u}{R_{AdS}}\right)^{2}\left[-dt^{2}+dx_{i}^{2}+\left(1-\left(\frac{u_{\Lambda}}{u}\right)^{n}\right)dx_{comp}^{2}\right]+\frac{1}{2\pi\alpha'}\left(\frac{R_{AdS}}{u}\right)^{2}\frac{du^{2}}{\left(1-\left(\frac{u_{\Lambda}}{u}\right)^{n}\right)}\\
i & =1,...,\left(n-2\right)\end{aligned}
\end{equation}
 where $u_{\Lambda}^{n}$ is the energy density on the brane. Therefore
the $f$ and $g$ functions take the following forms\begin{equation}
\begin{aligned}f^{2}\left(u\right) & =\frac{1}{\left(2\pi\alpha'\right)^{2}}\left(\frac{u}{R_{AdS}}\right)^{4}\\
g^{2}\left(u\right) & =\frac{1}{\left(2\pi\alpha'\right)^{2}}\frac{1}{1-\left(\frac{u_{\Lambda}}{u}\right)^{n}}\end{aligned}
\end{equation}
 This background satisfies both conditions for confinement at long
distances and $\frac{1}{\ell}$ potential at short distances, for
any value of $n$ (larger than 2). For these conditions we have to
add the demand that the effective boundary theory at long distances
would be four dimensional. Therefore, a good candidate would be the
non-critical $AdS_{6}$ which has a five dimensional boundary with
one compactified coordinate (and hence a four dimensional effective
field theory on the boundary, following a similar argument as in the
D branes discussion).

\subsubsection*{Klebanov Strassler\label{sub:Klebanov-Strassler}}

The supergravity solution of the deformed conifold is of the following
form \cite{key-13,key-12} \begin{equation}
ds^{2}=h^{-\frac{1}{2}}\left(\tau\right)dx_{0123}^{2}+h^{\frac{1}{2}}\left(\tau\right)ds_{6}^{2}\label{eq:Klebanov-Strassler-metric}\end{equation}
 where $ds_{6}^{2}$ is the metric of the deformed conifold\begin{equation}
ds_{6}^{2}=\frac{\epsilon^{4/3}}{2}K\left(\tau\right)\left[\frac{1}{3K^{3}\left(\tau\right)}\left(d\tau^{2}+\left(g^{5}\right)^{2}\right)+\cosh^{2}\left(\frac{\tau}{2}\right)\left[\left(g^{3}\right)^{2}+\left(g^{4}\right)^{2}\right]+\sinh^{2}\left(\frac{\tau}{2}\right)\left[\left(g^{1}\right)^{2}+\left(g^{2}\right)^{2}\right]\right]\end{equation}
 $\epsilon$ is the energy scale and the functions $h\left(\tau\right)$
and $K\left(\tau\right)$ are given by\begin{equation}
h\left(\tau\right)=\left(g_{s}M\alpha'\right)^{2}2^{2/3}\epsilon^{-8/3}I\left(\tau\right)\end{equation}
 \begin{equation}
I\left(\tau\right)\equiv\int_{\tau}^{\infty}dx\frac{x\coth x-1}{\sinh^{2}x}\left(\sinh2x-2x\right)^{\frac{1}{3}}\end{equation}
 \begin{equation}
K\left(\tau\right)=\frac{\left(\sinh\left(2\tau\right)-2\tau\right)^{\frac{1}{3}}}{2^{\frac{1}{3}}\sinh\tau}\end{equation}
 $\tau$ is the radial coordinate, it is a pure numer (i.e. unitless)
running from zero to infinity on the boundary. Near the boundary,
i.e. at large $\tau$, the $h\left(\tau\right)$ and $K\left(\tau\right)$
function takes the form\begin{equation}
h\left(\tau\gg1\right)=\left(g_{s}M\alpha'\right)^{2}3\cdot2^{2/3}\epsilon^{-8/3}\tau e^{-\frac{4}{3}\tau}\end{equation}
 \begin{equation}
K\left(\tau\gg1\right)=2^{\frac{1}{3}}e^{-\frac{1}{3}\tau}\end{equation}
 From the form of the metric we conclude that the $f$ and $g$ functions
are given by\begin{equation}
\begin{aligned}f^{2}\left(\tau\right) & =h^{-1}\left(\tau\right)\\
g^{2}\left(\tau\right) & =\frac{\epsilon^{4/3}}{6}\frac{1}{K^{2}\left(\tau\right)}\end{aligned}
\end{equation}

Following \cite{key-12} one can check that the KS background is confining
since $f\left(\tau\right)$ has a minimum at $\tau=0$ and it is finite
there $f\left(0\right)\neq0$. On the other hand, near the boundary
($\tau\gg0$) this backgroud do not follows the form \eqref{eq:ansatz.asy.Coulombic}
required for the analysis in section \eqref{sec:Asymptotically Coulombic}.
However, it is approximately $AdS$ in this region, therefore we expect
it to generate a potential which is approximately $\frac{1}{\ell}$
at short distances. Near the boundary the KS metric takes the form
\begin{equation}
ds^{2}\sim\epsilon^{4/3}\tau^{-\frac{1}{2}}e^{\frac{2}{3}\tau}dx_{0123}^{2}+\tau^{\frac{1}{2}}d\tau^{2}+\ldots\label{eq: near-boundary-KS-metric}\end{equation}
 The behavior of the wrap factors is dominated by the exponential
term. Therefore, by neglecting the power law terms of the wrap factors
we change the geometry just by little. The resultant approximate geometry
is then \begin{equation}
ds^{2}\sim\epsilon^{4/3}e^{\frac{2}{3}\tau}dx_{0123}^{2}+d\tau^{2}+\ldots\end{equation}
 which can be brought, by an appropriate coordinate redefinition,
to the form of an $AdS$ geometry\begin{equation}
ds^{2}\sim\left(\frac{u}{R}\right)^{2}dx_{0123}^{2}+\left(\frac{R}{u}\right)^{2}du^{2}+\ldots\end{equation}
 Hence, since the $AdS$ geometry generates a $\frac{1}{\ell}$ potential,
we expect the KS background to generate a potential which is approximately
$\frac{1}{\ell}$ near the boundary. Then the arising question is
how do the corrections to the approximate $\frac{1}{\ell}$ potential
look like? To answer this question we use the theorem reviewed in
\eqref{sec:The holographic Wilson line} (the exact analysis is detailed
in appendix \eqref{sec:The-potential-for-KS}). Plugging the discussed
geometry \eqref{eq: near-boundary-KS-metric} into the expression
for the length of the string \eqref{eq:length} one finds\begin{equation}
\ell\left(\tau_{0}\right)\sim\epsilon^{2/3}\frac{1}{f\left(\tau_{0}\right)}e^{+\frac{1}{3}\tau_{0}}\sim\frac{g_{s}M\alpha'}{\epsilon^{2/3}}\tau_{0}^{\frac{1}{2}}e^{-\frac{1}{3}\tau_{0}}\end{equation}
 In the same way the Kappa function \eqref{eq:Kappa} takes the form\begin{equation}
\mathcal{K}\left(\tau_{0}\right)\sim\epsilon^{2/3}e^{\frac{1}{3}\tau_{0}}\end{equation}
 Then the energy of the string \eqref{eq:Energy} would be\begin{equation}
E\left(\tau_{0}\right)\sim-\epsilon^{2/3}e^{\frac{1}{3}\tau_{0}}\end{equation}
 In order to derive the dependence of the energy on the length we
need to invert the function $\ell\left(\tau_{0}\right)$ into $\tau_{0}\left(\ell\right)$
and plug it into the expression for the energy $E\left(\ell\right)\sim-\epsilon^{2/3}e^{\frac{1}{3}\tau_{0}\left(\ell\right)}$.
This is very difficult to perform. Instead we will expand $\ell\left(\tau_{0}\right)$
around large $\tau_{0}$ and then invert the relation. For that it
would be usefull to take the logarithm of the length \begin{equation}
\ln\left[\left(g_{s}M\alpha'\right)^{-1}\epsilon^{2/3}\ell\right]\sim-\frac{1}{3}\tau_{0}+\ln\sqrt{\tau_{0}}\label{eq:KSsection.Length.Exact}\end{equation}
 The leading order\begin{equation}
\ln\left[\left(g_{s}M\alpha'\right)^{-1}\epsilon^{2/3}\ell\right]\sim-\frac{1}{3}\tau_{0}\label{eq:KSsection.Length.leading order}\end{equation}
 would result in the leading $\frac{1}{\ell}$ behavior\begin{equation}
E\sim-\frac{g_{s}M\alpha'}{\ell}\end{equation}
 as expected. In order to calculate the leading correction we would
plug the leading order relation \eqref{eq:KSsection.Length.leading order}
into the subleading term of the exact expression \eqref{eq:KSsection.Length.Exact}\begin{equation}
\ln\left[\left(g_{s}M\alpha'\right)^{-1}\epsilon^{2/3}\ell\right]\sim-\frac{1}{3}\tau_{0}+\ln\sqrt{-\ln\left[\left(g_{s}M\alpha'\right)^{-1}\epsilon^{2/3}\ell\right]}\end{equation}
 to get the dependence of $\tau_{0}$ on $\ell$\begin{equation}
\frac{1}{3}\tau_{0}\sim-\ln\left[\left(g_{s}M\alpha'\right)^{-1}\epsilon^{2/3}\ell\right]+\ln\sqrt{-\ln\left[\left(g_{s}M\alpha'\right)^{-1}\epsilon^{2/3}\ell\right]}\end{equation}
 Inserting this into the expression for the energy we finish with
a potential of the form\begin{equation}
E\left(\ell\right)\sim-\frac{\left(g_{s}M\alpha'\right)\sqrt{-\ln\left[\left(g_{s}M\alpha'\right)^{-1}\epsilon^{2/3}\ell\right]}}{\ell}\end{equation}
 In principle, there could appear also a constant term, as will be
discussed in section \eqref{sec:The-constant-terms-section}. In appendix
\eqref{sec:The-potential-for-KS} we explicitly calculate it.

\section{Units of measurement\label{sec:Units-of-measurement}}

In order to make a quantitative comparison between different models
one has to discuss the issue of units of measurement. In this section
we discuss two issues concerning this subject. We start with an abstract
discussion and then specialize to the models under investigation.

The first issue is the question of how to determine the system of
units when infinitely distant scales are introduced. In a theory with
only one scale parameter (say, the fundamental string length $l_{s}$)
every physical quantity $Q$ will be expressed as some power of the
only scale\begin{equation}
Q\sim\left(l_{s}\right)^{p}\end{equation}
 Now consider the case in which another scale $R$ is introduced.
When the two scales are very far apart $Q$ would have to be expressed
as some powers of both\begin{equation}
Q\sim\left(l_{s}\right)^{p_{1}}\left(R\right)^{p_{2}}\end{equation}
 If we further seperate the two scales by an infinite gap, there would
be a unique choice of the powers $p_{1}$ and $p_{2}$ such as to
get finite answers, and this is the way one has to choose his system
of units. Any other system would make no sense since it would result
with infinite values for physical quantities.

The second question is how to compare different models which have
different scale parameters. Each model has its own scale parameters
and therefore the immediate system of units is different in each of
them. In order to quantitatively compare different models one has
to describe the quantities under comparison in a model-independent
way, i.e. they have to be expressed in terms of universal physical
quantities that are present in all the models under investigation.

Specializing to the framework we explore in this paper, the physical
quantities we are interested in are the energy of the string and the
distance between its endpoints on the boundary, the scales are the
fundamental string length $l_{s}$ and the radius of the space $R_{space}$,
and the fact that the two scales are infinitely far away from each
other is exactly the statement of weakly curved holography\begin{equation}
\alpha'=l_{s}^{2}\ll R_{space}^{2}\end{equation}
 It is clear now that since $l_{s}$ and $R_{space}$ do not exist
in the lexicon of the gauge theory, we need to express the energy
and length in terms of universal physical quantities which are present
in both frameworks, stringy holography and gauge theory. Since we
are dealing with aspects of strong interaction, the natural universal
physical quantities are the string tension and the glueball mass (because
they must exist in any model that aims to describe strong interactions).
In this language, the SUGRA limit is\begin{equation}
\sigma\gg m_{glueball}^{2}\end{equation}
 Being expressed in such a way, we would be able to compare the Wilson
line of these two frameworks on equal footing.

As a realization of this general discussion, we first consider the
near extremal $D_{p}$ brane. In appendix \eqref{sec:Analytical-expressions-for-Dp brane}
we apply the theorem presented in section \eqref{sec:The holographic Wilson line}
to find the expressions for the energy and length of the string in
this case. We find that the energy of a classical string is well-defined
when it is measured in units of \begin{equation}
\left[E\right]=\frac{u_{\Lambda}}{\alpha'}\end{equation}
 where $u_{\Lambda}$ is the scale parameter and $\alpha'$ is related
to the fundamental string length. The length of a classical string
is well-defined when it is measured in units of \begin{equation}
\left[\ell\right]=u_{\Lambda}\left(\frac{R_{AdS}}{u_{\Lambda}}\right)^{\frac{7-p}{2}}\end{equation}
 where $R_{AdS}$ is the radius of the $AdS$ space and is of the
same order as $u_{\Lambda}$. In this system of units, following equation
\eqref{eq:string tension}, the string tension will take the value\[
\sigma=\frac{1}{2\pi}\]
 in units of (the units of the string tension are essentially the
units of the energy over the units of length)\[
\left[\sigma\right]=\frac{\left[E\right]}{\left[\ell\right]}=\frac{1}{\alpha'}\left(\frac{u_{\Lambda}}{R_{AdS}}\right)^{\frac{7-p}{2}}\]
 Any other choice of system would result in an infinite value for
the string tension in the SUGRA limit. The glueball mass scale is
related to the radius of the compactified coordinate\begin{equation}
m_{gb}\sim\frac{1}{R_{x_{compactified}}}\end{equation}
 which is uniquely determined from the demand of avoiding a conical
singularity. In appendix \eqref{sec:Conical-Singularity-in Dp brane}
we review this calculation and show that in the case of the near extremal
$D_{p}$ brane background it is given by \begin{equation}
2\pi R_{x_{compactified}}=\frac{4\pi}{7-p}u_{\Lambda}\left(\frac{u_{\Lambda}}{R_{AdS}}\right)^{-\frac{7-p}{2}}\end{equation}
 Then the glueball mass is \begin{equation}
m_{gb}\sim\frac{1}{u_{\Lambda}}\left(\frac{u_{\Lambda}}{R_{AdS}}\right)^{\frac{7-p}{2}}\label{eq:gb mass Dp brane}\end{equation}
 The relation between the string tension ($\sigma=\frac{1}{2\pi}\frac{1}{\alpha'}\left(\frac{u_{\Lambda}}{R_{AdS}}\right)^{\frac{7-p}{2}}$)
and the glueball mass would take the form \begin{equation}
m_{gb}\sim\frac{\alpha'}{u_{\Lambda}}\sigma\label{eq:mgb-sigma relation}\end{equation}
 Now we can express the energy and length in terms of $\sigma$ and
$m_{gb}$\begin{equation}
\left[E\right]=\frac{u_{\Lambda}}{\alpha'}\sim\frac{\sigma}{m_{gb}}\label{eq:Energy.UoM}\end{equation}
 \begin{equation}
\left[\ell\right]=u_{\Lambda}\left(\frac{R_{AdS}}{u_{\Lambda}}\right)^{\frac{7-p}{2}}\sim\frac{1}{m_{gb}}\label{eq:Length.UoM}\end{equation}
 Being expressed in such a way, the energy and length are well-defined
and their values on both sides (holography and gauge theory) are compareable
(given that they are expressed in a similar way in the gauge theory).
The exact translation is derived in the next section.

For the case of non-critical $AdS_{n+1}$ a similar analysis holds.
The string tension is given by\begin{equation}
\sigma=\frac{1}{2\pi\alpha'}\left(\frac{u_{\Lambda}}{R_{AdS}}\right)^{2}\end{equation}
 As the periodicity of the compactified dimension is given by $\beta\equiv2\pi R_{AdS}=\frac{4\pi}{n}\frac{R^{2}}{u_{\Lambda}}$
(see appendix \eqref{sec:Conical-Singularity-in Dp brane}), the glueball
mass would be\begin{equation}
m_{gb}\sim\frac{u_{\Lambda}}{R_{AdS}^{2}}\end{equation}
 and the same relation \eqref{eq:mgb-sigma relation} between $m_{gb}$
and $\sigma$ holds. Using the appropriate units for the energy and
length found in appendix \eqref{sec:Analytical-expressions-for-AdS}
we reconstruct \eqref{eq:Energy.UoM} and \eqref{eq:Length.UoM} for
the present case \begin{equation}
\left[E\right]=\frac{u_{\Lambda}}{\alpha'}\sim\frac{\sigma}{m_{gb}}\end{equation}
 \begin{equation}
\left[\ell\right]=\frac{R_{AdS}^{2}}{u_{\Lambda}}\sim\frac{1}{m_{gb}}\end{equation}

Next we consider the \textit{Klebanov Strassler} construction. The
string tension is given by \eqref{eq:string tension}\begin{equation}
\sigma=f\left(0\right)=h^{-\frac{1}{2}}\left(0\right)=2^{-1/3}I\left(0\right)\frac{\epsilon^{4/3}}{g_{s}M\alpha'}\cong0.57\frac{\epsilon^{4/3}}{g_{s}M\alpha'}\end{equation}
 and the glueball mass is \cite{key-13}\begin{equation}
m_{gb}\sim\frac{\epsilon^{2/3}}{g_{s}M\alpha'}\end{equation}
 where $\epsilon$ is the scale parameter and $g_{s}$ is the fundamental
string coupling and $M$ is the rank of the symmetry group. Using
the form of the metric \eqref{eq:Klebanov-Strassler-metric} one can
find the appropriate units for the energy and length of the string%
\footnote{Since the radial coordinate is dimensionless, the units of the energy
are essentially the units of the $g$ function whereas the units of
the length are the units of the $g$ function over the units of the
$f$ function.%
} and express them in terms of the string tension and the glueball
mass

\begin{equation}
\left[E\right]=\epsilon^{2/3}\sim\frac{\sigma}{m_{gb}}\end{equation}
 \begin{equation}
\left[\ell\right]=\frac{g_{s}M\alpha'}{\epsilon^{2/3}}\sim\frac{1}{m_{gb}}\end{equation}

The exact translations between the natural system of units of each
holographic model to the universal one is derived in the next section.

\section{Comparison with lattice gauge theories\label{sec:Comparison-with-lattice}}

In this section we will compare the potential of the holographic models
to lattice results of $SU\left(2\right)$ and $SU\left(3\right)$
gauge theories without fermions \cite{key-8,key-9}. In order to do
so we first have to express everything in terms of universal physical
quantities, as described in section \eqref{sec:Units-of-measurement}.
Our next task would be to determine the exact translation between
the natural system of units of each holographic model to the universal
one. In table \eqref{Table:String tensions and glueball masses in holography}
we summarize the results for the string tension and the lowest $0^{++}$
glueball mass%
\footnote{The reason for choosing the $0^{++}$ state for the universal system
of units is that in $SU\left(N\right)$ gauge theory it has the lowest
mass.%
} of the various holographic models. The lowest $0^{++}$ glueball
mass for the $D_{3}$, $D_{4}$, $AdS_{6}$ and $KS$ models was calculated
in \cite{Csaki:1998qr}, \cite{Minahan:1998tm}, \cite{Kuperstein:2004yf}
and \cite{Berg:2006xy}, respectively. The units of \cite{Berg:2006xy}'s
results are explicitly resolved in \cite{Benna:2007mb}. Then it is
simple to invert the relations such as to result with a dictionary
of translation between the natural system of units of each holographic
model to the universal one, see table \eqref{Table:Dictionary of translation}.

\begin{center}
\begin{table}
\begin{centering}
\begin{tabular}{|c|c|c|}
\hline
 & $\sigma$  & $m_{gb}=m_{0^{++}}$\tabularnewline
\hline
\hline
$D_{3}$ brane  & $\frac{1}{2\pi\alpha'}\left(\frac{u_{\Lambda}}{R_{AdS}}\right)^{2}$  & $\sqrt{11.59}\frac{u_{\Lambda}}{R_{AdS}^{2}}\cong3.4\frac{u_{\Lambda}}{R_{AdS}^{2}}$\tabularnewline
\hline
$D_{4}$ brane  & $\frac{1}{2\pi\alpha'}\left(\frac{u_{\Lambda}}{R_{AdS}}\right)^{\frac{3}{2}}$  & $\left(9.39\cdot\frac{3}{4\pi}\right)\frac{U_{\Lambda}^{1/2}}{R_{AdS}^{3/2}}\cong2.24\frac{U_{\Lambda}^{1/2}}{R_{AdS}^{3/2}}$\tabularnewline
\hline
$AdS_{6}$  & $\frac{1}{2\pi\alpha'}\left(\frac{u_{\Lambda}}{R_{AdS}}\right)^{2}$  & $\left(6.34\cdot\frac{5}{4\pi}\right)\frac{u_{\Lambda}}{R_{AdS}^{2}}\cong2.52\frac{u_{\Lambda}}{R_{AdS}^{2}}$\tabularnewline
\hline
$KS$  & $0.57\frac{\epsilon^{4/3}}{g_{s}M\alpha'}$  & $0.86\frac{\epsilon^{2/3}}{g_{s}M\alpha'}$\tabularnewline
\hline
\end{tabular}
\par\end{centering}

\caption{\label{Table:String tensions and glueball masses in holography}String
tension and lowest $0^{++}$ glueball mass of the various holographic
models.}

\end{table}

\par\end{center}

\begin{center}
\begin{table}
\begin{centering}
\begin{tabular}{|c|c|c|}
\hline
 & $\left[E\right]$  & $\left[\ell\right]$\tabularnewline
\hline
\hline
$D_{3}$ brane  & $\frac{u_{\Lambda}}{\alpha'}=\left(3.4\cdot2\pi\right)\frac{\sigma}{m_{gb}}$  & $\frac{R_{AdS}^{2}}{u_{\Lambda}}=3.4\frac{1}{m_{gb}}$\tabularnewline
\hline
$D_{4}$ brane  & $\frac{u_{\Lambda}}{\alpha'}=\left(2.24\cdot2\pi\right)\frac{\sigma}{m_{gb}}$  & $\frac{R_{AdS}^{3/2}}{u_{\Lambda}^{1/2}}=2.24\frac{1}{m_{gb}}$\tabularnewline
\hline
$AdS_{6}$  & $\frac{u_{\Lambda}}{\alpha'}=\left(2.52\cdot2\pi\right)\frac{\sigma}{m_{gb}}$  & $\frac{R_{AdS}^{2}}{u_{\Lambda}}=2.52\frac{1}{m_{gb}}$\tabularnewline
\hline
$KS$  & $\epsilon^{2/3}=1.51\frac{\sigma}{m_{gb}}$  & $\frac{g_{s}M\alpha'}{\epsilon^{2/3}}=0.86\frac{1}{m_{gb}}$\tabularnewline
\hline
\end{tabular}
\par\end{centering}

\caption{\label{Table:Dictionary of translation}Dictionary of translation
between the natural system of units of each holographic model to the
universal one.}

\end{table}

\par\end{center}

The next step would be to express the system of units used by the
lattice conventions in terms of the universal physical quantities
discussed above. In the lattice conventions, length is measured in
units $r_{0}$ related to the string tension via the next expression
\cite{key-9}\begin{equation}
\sigma r_{0}^{2}=1.65-\frac{\pi}{12}\end{equation}
 The value of the lowest $0^{++}$ glueball mass in $D=3+1$ dimensions
is given by \cite{Teper:1998te}\begin{equation}
m_{glueball}\equiv m_{0^{++}}=d_{N}\cdot\sqrt{\sigma}\end{equation}
\[
d_{N}=\begin{cases}
3.87 & N=2\\
3.65 & N=3\end{cases}\]
Hence we can deduce the relation between the original lattice length's
unit to the desired ones\begin{equation}
r_{0}=d_{N}\sqrt{1.65-\frac{\pi}{12}}\frac{1}{m_{gb}}\end{equation}
 In the same manner the translation of the energy units, which are
given in the lattice convention by $\frac{1}{r_{0}}$, is done using
the next expression\begin{equation}
\frac{1}{r_{0}}=\frac{d_{N}}{\sqrt{1.65-\frac{\pi}{12}}}\frac{\sigma}{m_{gb}}\end{equation}

The exact expressions for the potentials in $D_{p}$ brane, non-critical
$AdS_{n+1}$ and \textit{Klebanov-Strassler} backgrounds are derived
explicilty in appendices \eqref{sec:Analytical-expressions-for-Dp brane},
\eqref{sec:Analytical-expressions-for-AdS} and \eqref{sec:The-potential-for-KS},
respectively. They are presented together with the lattice data in
terms of the universal system of units and using two different conventions
\begin{enumerate}
\item Fixing the zero energy point (as in the conventions of \cite{key-9}).
\item Comparing the confining asymptotic behavior of all the graphs.
\end{enumerate}
in figures \eqref{Fig:Potential.Graph.FixedZeroEnergy} and \eqref{Fig:Potential.Graph.UnifiedCA},
respectively.

\clearpage
\newpage

\begin{center}
\begin{figure}
\begin{centering}
\includegraphics[scale=0.6]{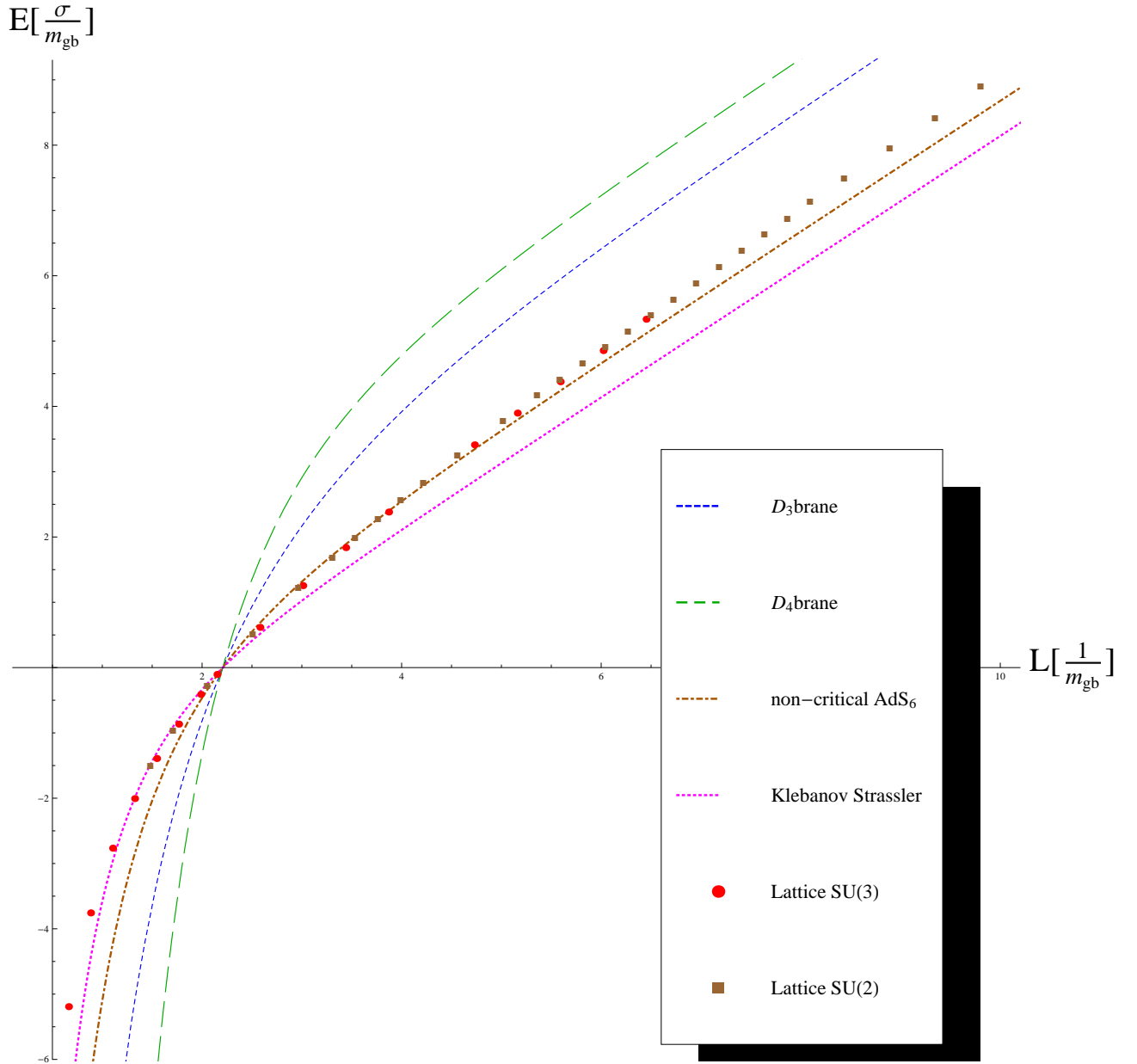}
\par\end{centering}

\caption{\label{Fig:Potential.Graph.FixedZeroEnergy}The heavy quarks potential
of the holographic models and the lattice data expressed in the appropriate
physical units and presented using the fixed zero energy point convention.}

\end{figure}

\par\end{center}

\clearpage
\newpage

\begin{center}
\begin{figure}
\begin{centering}
\includegraphics[scale=0.6]{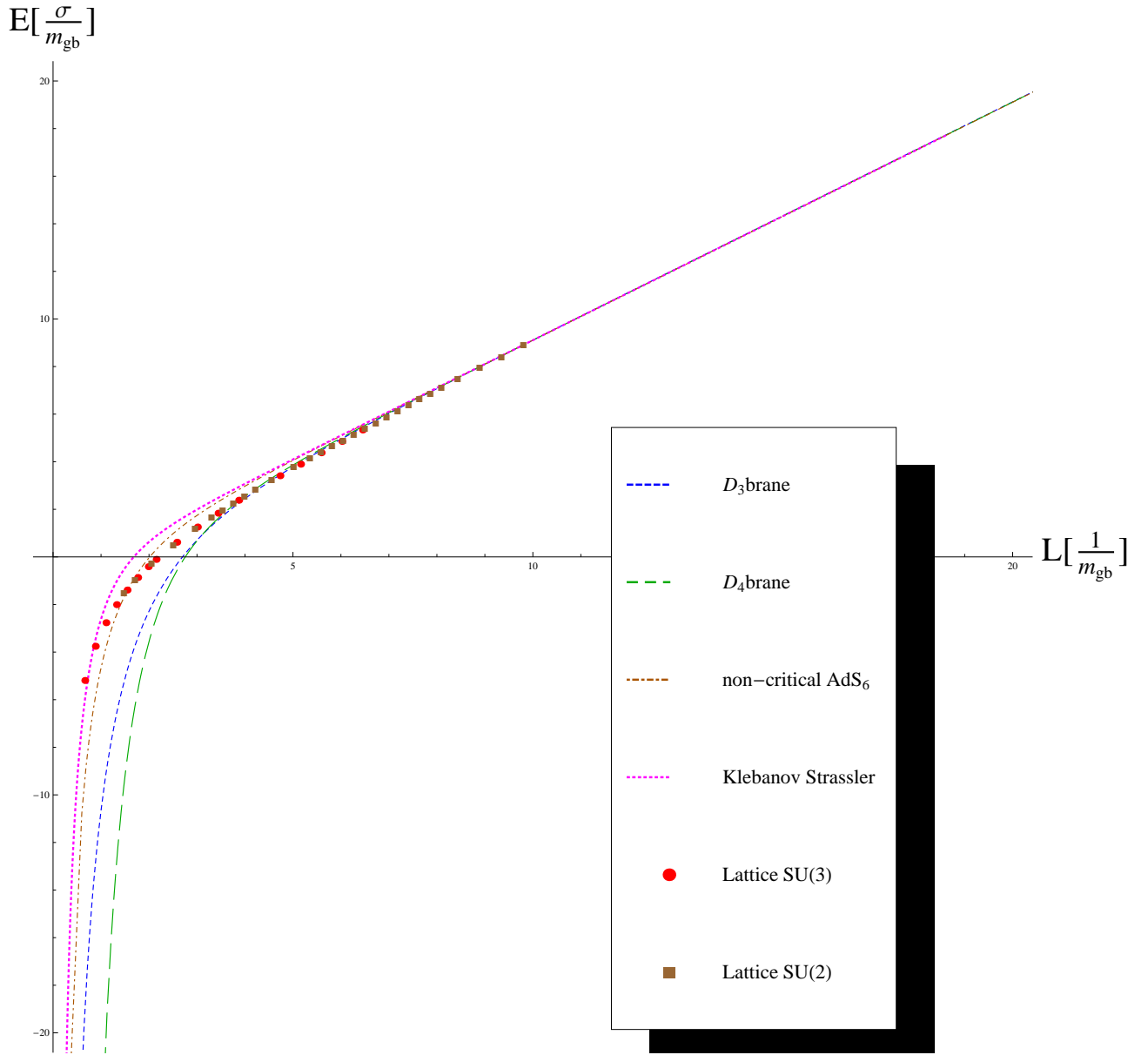}
\par\end{centering}

\caption{\label{Fig:Potential.Graph.UnifiedCA}The heavy quarks potential of
the holographic models and the lattice data expressed in the appropriate
physical units and presented using the convention in which we compare
the confining asymptotic behavior of all the graphs.}

\end{figure}

\par\end{center}

\clearpage
\newpage
In order to quantify the measure of correspondence of the holographic
models' potential to the lattice data we will use Pearson's $\chi^{2}$
test\begin{equation}
\chi^{2}=\Sigma_{i=1}^{n}\frac{\left(O_{i}-E_{i}\right)^{2}}{|E_{i}|}\end{equation}
 where $n$ is the number of the lattice data points. The computation
is done using the two conventions described above and in addition
we present another convention in which the height of the theoretical
potential curve is adjusted to result with a minimal $\chi^{2}$,
i.e. the best fit possible. We calculate $\chi^{2}$ only for the
correspondence with the SU(3) simulation%
\footnote{We read \cite{key-8}'s results for the SU(2) simulation out of the
graph since they are not given explicitly, therefore they will be
used only for a qualitative picture and not for quantitive measures%
}. The results are presented in table \eqref{Table:chi^2 table}.

\begin{center}
\begin{table}
\begin{centering}
\begin{tabular}{|c|c|c|c|}
\hline
~~~~~~~~~~~~~Convention$\rightarrow$  & Fixed zero  & Best fit & Unified confining \tabularnewline
Case$\downarrow$~~~~~~~~~~~~~~~~~~~~~~~  & energy point  &  & asymptotic behavior \tabularnewline
\hline
\hline
near extremal $D_{3}$ brane  & 24.6 & 24.4 & 34.8\tabularnewline
\hline
near extremal $D_{4}$ brane  & 108 & 108 & 120\tabularnewline
\hline
non critical $AdS_{6}$  & 4.14 & 3.98 & 5.89\tabularnewline
\hline
Klebanov Strassler  & 1.4 & 1.33 & 14.5\tabularnewline
\hline
\end{tabular}
\par\end{centering}

\caption{\label{Table:chi^2 table}$\chi^{2}$ values for the measure of correspondence
between the holographic models' potential and the lattice measurements
for SU(3). The calculation was done using three different conventions,
as detailed in the text.}

\end{table}

\par\end{center}

From the results we can determine that two of the models we have examined
correspond significantly better to the lattice measurement of the
Wilson line than the two other ones. The good models are the \textit{Klebanov
Strassler} and the non-critical $AdS_{6}$ backgrounds which have
significantly much better $\chi^{2}$ values than the near extremal
D branes. On the other hand, the $D_{4}$ brane background generates
a potential which do not fit at all to the lattice potential, with
$\chi^{2}$'s greater than 100 for all the conventions of comparison
we have used.

Out of this summary we can already conclude that the requirement for
a Coulombic behavior at short distances proves itself in the sense
that the models which admit at short distances a potential which is
either exactly or approximately of the form $\sim\frac{1}{\ell}$
are also in a good agreement with the lattice measurements, compared
to the $D_{4}$ brane which admits a $\sim\frac{1}{\ell^{2}}$ potential
in this regime and has a $\chi^{2}$ value which is far above all
the other. Hence it makes sense to use this requirement when searching
for an holographic model which mimics QCD's Wilson line.

The \textit{best fit} convention of comparison gives almost identical
results to the \textit{fixed zero energy point} convention. In both,
the model which has the best correspondence with the lattice results
is the $KS$ with a remarkable $\chi$ square value, but nevertheless
the $AdS_{6}$ also fits quite well. In the third convention (\textit{unified
confining asymptotic behavior}) they flip over as the $KS$ becomes
secondary to the $AdS_{6}$ which best fits.

\section{The constant terms\label{sec:The-constant-terms-section}}

At the beginning of this paper we have made a statement that at large
distances the potential is linear and at small distances it is Coulombic.
However, there is still a possibility of an additional constant term
for each of these asymptotic behaviors, i.e.\begin{eqnarray}
E_{long} & = & \sigma\ell+C_{L}\\
E_{short} & \sim & -\frac{1}{\ell}+C_{S}\end{eqnarray}
 In a sense, these constant terms characterize the behavior of the
potential at intermediate distances. Therefore, a comparison between
such constants of the holographic models to the constants of the gauge
theory would constitute a measure for the similarity of the potentials
at intermediates.

In principle, each of the constants ($C_{L}$ and $C_{S}$) for its
own is meaningless and only their difference\begin{equation}
\Delta C\equiv C_{L}-C_{S}\end{equation}
 is meaningful, since the energy can always be shifted by a constant.
Therefore one should expect to consider only their difference. However,
since we have fixed the zero energy point there is still a possibility
to look at each of the constants separately. Of course, that would
depend on the way we have fixed the energy, but we will still find
it usefull to compare the large distance constant of the different
models and the lattice, as will be described later.

Next, we will check whether such constants do appear in the holographic
models. In appendices \eqref{sec:The-values-of-constants-DpBranes},\eqref{sec:The-value-of-the-constant-AdS}
and \eqref{sec:The-potential-for-KS} we calculate the values of $C_{L}$
and $C_{S}$ for the different cases, before fixing the zero energy
point as in \cite{key-9}'s notations. The results we get \textit{after}
fixing the zero energy point are listed in table \eqref{Table:constant.terms.holo}.
Of course, the results for $\Delta C$ do not depend on how we have
fixed the zero point energy, therefore they are explained by the mentioned
appendices.

\begin{center}
\begin{table}
\begin{centering}
\begin{tabular}{|c|c|c|c|}
\hline
 & $C_{L}\left[\frac{\sigma}{m_{gb}}\right]$  & $C_{S}\left[\frac{\sigma}{m_{gb}}\right]$  & $\Delta C\left[\frac{\sigma}{m_{gb}}\right]$\tabularnewline
\hline
\hline
$D_{3}$ brane  & 0.59 & 7.39 & -6.8\tabularnewline
\hline
$D_{4}$ brane  & 1.35 & 5.3 & -4.48\tabularnewline
\hline
non critical $AdS_{6}$  & -1.32 & 4.02 & -5.34\tabularnewline
\hline
Klebanov Strassler  & -1.44 & 2.93 & -4.37\tabularnewline
\hline
\end{tabular}
\par\end{centering}

\caption{\label{Table:constant.terms.holo}The values of the constant terms
for the different models in the fixed zero energy point convention.}

\end{table}

\par\end{center}

From these results we can already conclude about the constant term
in holography by itself. The values of $\Delta C$ in all the models
under investigation are all non-zero, negative and very close to each
other ($\Delta C\simeq-5\frac{\sigma}{m_{gb}}$), such that several
remarks are to be made:
\begin{itemize}
\item It seems that generically holography generates such a constant term.
We can get some intuition about this feature of holography by thinking
of how one would generate it in a flat space-time Nambu-Goto theory.
While at short distances a tachyonic instability takes over such that
there is no sense in talking about a constant term in this region,
at long distances it does make sense, but as happens - it does not
shows up (see eq. \eqref{NG}). The only way of generating such a
term at long distances is by adding a boundary term to the Nambu-Goto
action. Therefore, since holoraphic models have a boundary by definition,
we can expect them to have a constant term of the discussed type.
\item We interpret $\left(-\Delta C\right)$ as the physical mass of the
quarks which remains after subtracting their bare mass. Therefore
it is crucial that the values of $\Delta C$ in all the models have
the same sign.
\item The fact that all the holographic models admits a similar $\Delta C$
value is very interesting.
\end{itemize}
Now, we will derive the values of the constant terms for the gauge
theory, using the lattice results. By adjusting the theoretical prediction\begin{equation}
E=\sigma\ell-\frac{\pi}{12}\frac{1}{\ell}+C_{L}\end{equation}
 (where the first term is just the linear potential and the second
one is the well-known Luscher term) to the measurements, we find $C_{L}$.
Finding $C_{S}$ appears to be a subtle issue. In principle, one should
compare the theoretical prediction of perturbation theory to the measurements,
and derive $C_{S}$ in the same manner as for $C_{L}$. However, as
was found in \cite{key-10}, the pertubative expression of the potential
(up to three loops) fails in describing the potential even at the
shortest lengths reached by the lattice simulations. Therefore, by
adjusting this theoretical prediction to the shortest lattice measurement,
we can only get a rough estimate for $C_{S}$. We are required to
this point in the next paragraph, but first let us present the results,
listed in table \eqref{Table:constant.terms.lattice}. We use theoretical
predictions of 0,1 and 2 loop orders of perturbation theory, in order
to see how much $C_{L}$ changes as we go to high order expressions.

\begin{center}
\begin{table}
\begin{centering}
\begin{tabular}{|c|c|c|c|}
\hline
 & $C_{L}\left[\frac{\sigma}{m_{gb}}\right]$  & $C_{S}\left[\frac{\sigma}{m_{gb}}\right]$  & $\Delta C\left[\frac{\sigma}{m_{gb}}\right]$\tabularnewline
\hline
\hline
Lattice SU(3) 0-loop  & -0.63 & -2.24 & 1.61\tabularnewline
\hline
Lattice SU(3) 1-loop  & -0.63 & 6.31 & -6.94\tabularnewline
\hline
Lattice SU(3) 2-loop  & -0.63 & 2.48 & -3.11\tabularnewline
\hline
Lattice SU(2) 0-loop  & -0.43 & -0.01 & -0.42\tabularnewline
\hline
Lattice SU(2) 1-loop  & -0.43 & 11.67 & -12.1\tabularnewline
\hline
Lattice SU(2) 2-loop  & -0.43 & 5.94 & -6.37\tabularnewline
\hline
\end{tabular}
\par\end{centering}

\caption{\label{Table:constant.terms.lattice}The values of the constant terms
for SU(2) and SU(3) gauge theories. We used the data of \cite{key-8,key-9}
to calculate them.}

\end{table}

\par\end{center}

\begin{center}
\begin{figure}
\begin{centering}
\includegraphics[scale=0.8]{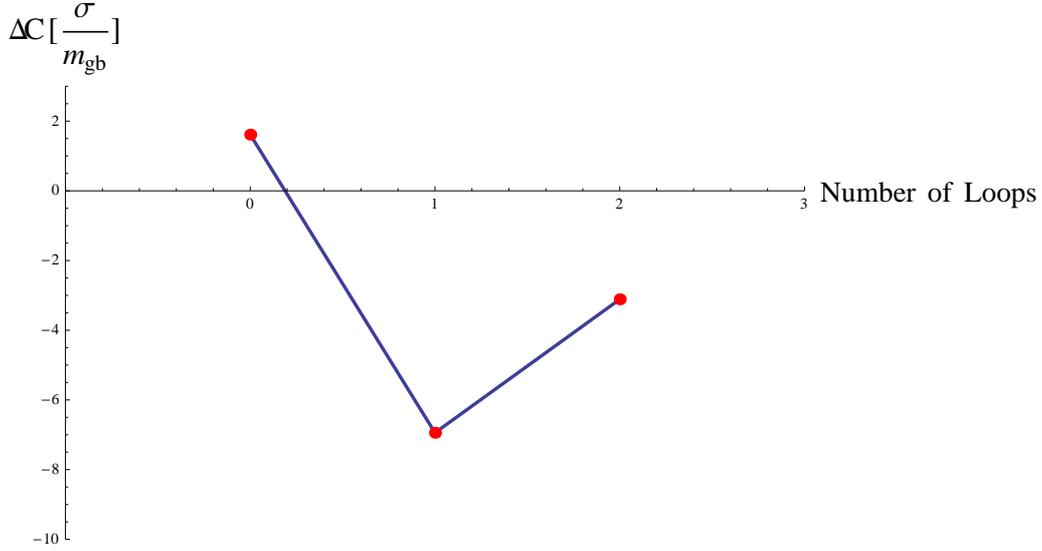}
\par\end{centering}

\caption{\label{Fig: const/No.ofLoops}$\Delta C$ values as were calculated
from comparison of the SU(3) lattice measurements to the theoretical
predictions of 0,1 and 2 loops orders of perturbation theory.}

\end{figure}

\par\end{center}

As was discussed above, the process of finding the constant term at
short distances (and therefore also $\Delta C$) out of the lattice
measurements is a subtle issue. One way of resolving this issue is
to consider only the constant at long distances, which is well defined
since we have fixed the zero energy point, but also somewhat arbitrary
defined, since this fixing is arbitrary. By examining the values of
the constant terms at long distances one can notice that all the holographic
models has $C_{L}$'s which are of the same order as the one extracted
out of the lattice data for the gauge theories. Moreover, our two
favored candidates ($KS$ \& $AdS_{6}$) have negative $C_{L}$ as
in the gauge theories (while the D branes have positive $C_{L}$),
which may be regarded as another support in them. We emphasize that
this conclusion should be taken with a grain of salt since it depends
on how we have fixed the zero energy point, but it is still interesting.
The other approach we take here is to look on the values of $\Delta C$
as were calculated from adjusting the theoretical predictions of 0,1
and 2 loops orders of perturbation theory to the lattice measurements
and try to estimate to which value it would converge if we were able
to use the theoretical expression of higher orders. From the results
of the SU(3) measurements (plotted in figure \eqref{Fig: const/No.ofLoops})
the guess is that $\Delta C$ would converge to a value somewhat around
$\Delta C=-5\frac{\sigma}{m_{gb}}$. Considering this value as a basis
for comparison, all the holographic models under investigation definitely
have a very similar $\Delta C$ value.

\section{Concluding remarks \& future directions\label{sec:Concluding-remarks}}

Following the statement made in the introduction, the goal of this
paper is to search for a holographic background that admits a Wilson
line which is as close as possible to the potential deduced from lattice
calculations and which at the same time is in accordance with other
requirements: Confinement at long distances, Coulombic potential at
short distances (by which we mean $\frac{1}{\ell}$ behavior)%
\footnote{This requirement is motivated by the fact that the leading perturbative
behavior of the QCD Wilson line is Coulombic.%
} and invariance under four dimensional Poincare symmetry.

We summarize by suggesting the \textit{Klebanov Strassler} background
as the holographic model which best fits the lattice measurements
and in the same time is in accordance with the other requirements
we have mentioned above\footnote{The KS potential at small distances is not exactly but rather approximately of the form $\sim\frac{1}{\ell}$.}. Nevertheless, the non-critical $AdS_{6}$
also has an appreciable correspondence and it also admits the other
requirements, so we cannot rule it out. However, from this point of
view, the near extremal D branes can undoubtedly be excluded.

We turn now to final remarks and future directions. The first remark
concerns the rank of the symmetry group, N. So far, the statement
of holography (or the gauge/gravity duality) is made about dualities
between string theories and field theories in the limit of large N.
On the other hand, we have used lattice simulations of SU(N) gauge
theories with finite N (N=2,3). Therefore one may argue that the comparison
we have made is not valid. However, the claim is that the discussed
Wilson line does not depends strongly on N. This claim is based on
results from recent lattice simulations of Wilson lines \cite{Athenodorou:2007du,Athenodorou:2010cs,Brandt:2010bw}
for several values of N. It can also be seen in the lattice results
we've been using (plotted in figure \eqref{Fig:Potential.Graph.FixedZeroEnergy}),
where there is no much difference between the SU(2) and SU(3) potentials.
Assuming this claim to be correct, the large N Wilson line of the
gauge theory should be similar to the ones we have used for the comparison.

The second remark concerns quantum corrections. As the analysis in
the present work is only at the classical level, the arising question
is how quantum corrections would affect the results. In order to understand
this point, we first review the situation in the flat space-time N.G.
case, there the potential at the classical level is just a linear
piece\[
E=\sigma\ell\]
 at all length scales. The full quantum expression \eqref{NG} presents
a dramatic change of this classical picture. While at long distances
the quantum corrections are suppressed so the classical expression
is a good approximation, at short length scales a tachyonic behavior
takes over such that beyond some critical length \eqref{NG.critical.length}
the full quantum expression \eqref{NG} is not valid anymore. In the
case of an asymptotically AdS curved background we expect the effect
of the quantum corrections to be less dramatic, as we will explain
below%
\footnote{However, there is an important difference between the cases. While
the tachyonic behavior in the case of flat space-time NG is an indication
for an instability of the theory, in the present construction of the
curved space-time case it is an indication for an instability of the
specific Wilson line we chose and not of the theory. The reason for
this is that in the latter case we have constrained the string's endpoints
to end on the boundary while in the former there are still dynamical
open strings. Therefore, by referring here to instability of the curved
background case we only mean instability of the Wilson line and not
of the theory. We leave this issue for future work.%
}. In the classical level, the potential is linear at long distances
but it becomes Coulombic ($\sim\frac{1}{\ell}$) at short distances.
The quantum corrections at long distances were proven to be suppressed;
in \cite{key-25} (\cite{key-26}, \cite{Aharony:2010db}) the effective
action of closed (open) confining strings was found to obey the Nambu-Goto
form up to high orders, i.e. deviations from the N.G. energy levels
are suppressed as some high power of $\frac{1}{\ell}$ (see also \cite{key-24}
where a discussion about quantum fluctuations of Wilson loop in string
models is made). On dimensional grounds, the quantum corrections at
short distances should have a Coulombic form (since the theory is
asymptotically conformal) such that the potential would take the form\begin{eqnarray*}
E & = & -\frac{4\pi^{2}}{\Gamma^{4}\left(\frac{1}{4}\right)}\frac{\sqrt{\lambda}}{\ell}\left[1+\frac{\tilde{\kappa}}{\sqrt{\lambda}}+\mathcal{O}\left(\frac{1}{\lambda}\right)\right]\\
\sqrt{\lambda} & \sim & \frac{1}{\alpha'}\end{eqnarray*}
 The one loop correction to the classical term was calculated in \cite{Drukker:2000ep}
and the value of $\tilde{\kappa}$ was evaluated numerically in \cite{key-23}%
\footnote{We thank Nadav Drukker for referring us to this paper.%
} and analytically in \cite{Forini:2010ek} to be\[
\tilde{\kappa}\simeq-1.33460\]
 Therefore, the quantum corrections are suppressed in the limit where
$\lambda\rightarrow\infty$ so at short enough distances the classical
expression will be dominant and no tachyonic behavior would pop out.
We stress that this is true only at large $\lambda$ (or equivalently
weakly curved background) which is not necessarily the case in QCD.
To emphasis the difference between the flat and curved space-time
we remind that in the former case the expansion of the Wilson line
is done in powers of $\frac{\alpha'}{\ell^{2}}$ while in the latter
the expansion is in powers of $\frac{1}{\sqrt{\lambda}}$ (or equivalently
in powers of $\alpha'$) due to the conformal symmetry. Hence, in
the flat space-time scenario the quantum corrections at some given
$\alpha'$ would always become dominant at short enough distances
and may present a tachyonic behavior, while in the asymptotically
conformal curved space-time they are controllable. From this point
of view we would prefer to consider the non-critical $AdS_{6}$ background
rather than the $KS$ since the former is asymptotically AdS and the
latter is not.

Concerning the issue of quantum corrections, some future directions
are proposed:
\begin{itemize}
\item Following the statement made above, in asymptotically AdS backgrounds
tachyonic behavior can arise only at intermediate distances, as opposed
to the flat space-time case in which the tachyonic behavior is present
below some critical length ($\ell<\ell_{cr}$). It would be interesting
to characterize this finite range of tachyonic behavior at intermediate
distances and in particular try to formulate conditions under which
it is going to zero, such that no tachyons are involved.
\item Generalizing the quantum analysis to backgrounds which are not asymptotically
AdS (like the $D_{p\neq3}$ branes).
\end{itemize}

\acknowledgments

We are very grateful to Ofer Aharony for many useful discussions and
for his critical comments on the manuscript. We would also like to
thank Barak Bringoltz, Nadav Drukker and Herbert Neuberger for fruitful
conversations. U.K. would like to thank Zohar Komargodski and Andrew
Zayakin for helpful discussions during the Cargèse Summer School 2010
and to Dmitry Melnikov for assistance with the Klebanov-Strassler
background. U.K. would like to dedicate a special thank to Shimon
Yankielowicz for advising, teaching and supporting.

This work was supported in part by a centre of excellence supported
by the Israel Science Foundation (grant number 1468/06), by a grant
(DIP H52) of the German Israel Project Cooperation, by a BSF grant,
by the European Network MRTN-CT-2004-512194 and by European Union
Excellence Grant MEXT-CT-2003-509661.

%%%%%%%%%%%%%%%%%%%%%%%%%%%%%%%%%%%%%%%%%%%%%%%%%%%%

\appendix
%dummy comment inserted by tex2lyx to ensure that this paragraph is not empty
%dummy comment inserted by tex2lyx to ensure that this paragraph is not empty
%dummy comment inserted by tex2lyx to ensure that this paragraph is not empty

\section{The conditions on the background to satisfy a coulombic potential
in the UV\label{sec:asymptotically coulombian appendix}}

This derivation is being done using notations and symbols of \cite{key-1}.
First we expand the $f,g$ functions around $u=\infty$\begin{equation}
\begin{aligned}f\left(u\right) & =a_{k}u^{k}+\mathcal{O}\left(u^{k-1}\right)\\
g\left(u\right) & =g\left(\infty\right)+b_{j}u^{-j}+\mathcal{O}\left(u^{-j+1}\right)\\
k,j & >0\end{aligned}
\end{equation}
 such that we can calculate the energy in the limit $u_{0}\gg u_{\Lambda}$
using these expansions. The length of the string is given by

\begin{equation}
\begin{aligned}\ell\left(u_{0}\right) & =2f\left(u_{0}\right)\int_{u_{0}}^{\infty}du\frac{g\left(u\right)}{f^{2}\left(u\right)}\frac{1}{\sqrt{1-\frac{f^{2}\left(u_{0}\right)}{f^{2}\left(u\right)}}}\\
 & =2a_{k}u_{0}^{k}\int_{u_{0}}^{\infty}du\frac{g\left(\infty\right)+b_{j}u^{-j}}{a_{k}^{2}u^{2k}}\frac{1}{\sqrt{1-\left(\frac{u_{0}}{u}\right)^{2k}}}\\
 & =2a_{k}u_{0}^{k}\left[\frac{g\left(\infty\right)}{a_{k}^{2}}\int_{u_{0}}^{\infty}du\frac{1}{u^{2k}}\frac{1}{\sqrt{1-\left(\frac{u_{0}}{u}\right)^{2k}}}+\frac{b_{j}}{a_{k}^{2}}\int_{u_{0}}^{\infty}du\frac{1}{u^{2k+j}}\frac{1}{\sqrt{1-\left(\frac{u_{0}}{u}\right)^{2k}}}\right]\\
 & =2u_{0}^{k}\left[\frac{g\left(\infty\right)}{a_{k}}\frac{1}{\left(u_{0}\right)^{2k-1}}\int_{1}^{\infty}dy\frac{1}{\left(y\right)^{2k}}\frac{1}{\sqrt{1-y^{-2k}}}+\frac{b_{j}}{a_{k}}\frac{1}{\left(u_{0}\right)^{2k+j-1}}\int_{1}^{\infty}d\frac{1}{y^{2k+j}}\frac{1}{\sqrt{1-y^{-2k}}}\right]\\
 & =\frac{2g\left(\infty\right)}{a_{k}}\frac{1}{\left(u_{0}\right)^{k-1}}C_{2k,2k}\left(\infty\right)+\frac{2b_{j}}{a_{k}}\frac{1}{\left(u_{0}\right)^{k+j-1}}C_{2k+j,2k}\left(\infty\right)\\
 & =\left(\frac{2g\left(\infty\right)}{a_{k}}C_{2k,2k}\left(\infty\right)\right)\frac{1}{\left(u_{0}\right)^{k-1}}+\mathcal{O}\left(\frac{1}{\left(u_{0}\right)^{k+j-1}}\right)\end{aligned}
\end{equation}
For the next steps we will need to use the invert function $u_{0}\left(\ell\right)$\begin{equation}
u_{0}\left(\ell\right)=\left(\frac{2g\left(\infty\right)}{a_{k}}C_{2k,2k}\left(\infty\right)\frac{1}{\ell}\right)^{\frac{1}{k-1}}\label{eq:invert.function}\end{equation}
The $\mathcal{K}$ function is given by\begin{equation}
\begin{aligned}\mathcal{K}\left(u_{0}\right) & =\int_{u_{0}}^{\infty}du\, g\left(u\right)\left[1-\sqrt{1-\frac{f^{2}\left(u_{0}\right)}{f^{2}\left(u\right)}}\right]+\int_{0}^{u_{0}}du\, g\left(u\right)\\
 & =\int_{u_{0}}^{\infty}du\,\left(g\left(\infty\right)+b_{j}u^{-j}\right)\left[1-\sqrt{1-\left(\frac{u_{0}}{u}\right)^{2k}}\right]+\int_{0}^{u_{0}}du\,\left(g\left(\infty\right)+b_{j}u^{-j}\right)\\
 & =\int_{1}^{\infty}dy\, u_{0}\left(g\left(\infty\right)+b_{j}y^{-j}u_{0}^{-j}\right)\left[1-\sqrt{1-y^{-2k}}\right]+\int_{0}^{1}du\, u_{0}\left(g\left(\infty\right)+b_{j}y^{-j}u_{0}^{-j}\right)\\
 & =\frac{1}{2}u_{0}g\left(\infty\right)C'_{2k,2k}\left(\infty\right)+\frac{1}{2}u_{0}^{1-j}b_{j}C'_{2k+j,2k}\left(\infty\right)\\
 & +u_{0}g\left(\infty\right)+\frac{b_{j}}{1-j}u_{0}^{1-j}\\
 & =u_{0}g\left(\infty\right)\left[1+\frac{1}{2}C'_{2k,2k}\left(\infty\right)\right]+u_{0}^{1-j}b_{j}\left[\frac{1}{1-j}+\frac{1}{2}C'_{2k+j,2k}\left(\infty\right)\right]\\
 & =u_{0}g\left(\infty\right)\left[1+\frac{1}{2}C'_{2k,2k}\left(\infty\right)\right]+\mathcal{O}\left(u_{0}^{1-j}\right)\end{aligned}
\end{equation}
Using \eqref{eq:invert.function} we can express it as a function
of $\ell$\begin{equation}
\begin{aligned}\mathcal{K}\left(\ell\right) & =\left\{ g\left(\infty\right)\left[1+\frac{1}{2}C'_{2k,2k}\left(\infty\right)\right]\left(\frac{2g\left(\infty\right)}{a_{k}}C_{2k,2k}\left(\infty\right)\right)^{\frac{1}{k-1}}\right\} \left(\frac{1}{\ell}\right)^{\frac{1}{k-1}}\end{aligned}
\end{equation}
Finally, the energy of the string as a function of its length would
be given by\begin{equation}
\begin{aligned}E\left(\ell\right) & =f\left(u_{0}\left(\ell\right)\right)\cdot\ell-2\mathcal{K}\left(\ell\right)\\
 & =\left\{ a_{k}\left(\frac{2g\left(\infty\right)}{a_{k}}C_{2k,2k}\left(\infty\right)\right)^{\frac{k}{k-1}}+g\left(\infty\right)\left[1+\frac{1}{2}C'_{2k,2k}\left(\infty\right)\right]\left(\frac{2g\left(\infty\right)}{a_{k}}C_{2k,2k}\left(\infty\right)\right)^{\frac{1}{k-1}}\right\} \left(\frac{1}{\ell}\right)^{\frac{1}{k-1}}\end{aligned}
\end{equation}
The conclusion is that in order to have a coulombic behaviour at small
distances we have to demand the following asymptotic form of $f$\begin{equation}
f\left(u\gg u_{\Lambda}\right)\sim u^{2}\end{equation}

\section{Conical singularity in near extremal $D_{p}$ brane\label{sec:Conical-Singularity-in Dp brane}}

We review the calculation of the periodicity of the compactified coordinate
in the case of near extremal $D_{p}$ brane background. The part of
the metric which includes the radial coordiante, $u$, and the compactified
one, $x$, is given by

\begin{equation}
ds^{2}\sim-\left(\frac{u}{R}\right)^{\frac{7-p}{2}}\left[1-\left(\frac{u_{\Lambda}}{u}\right)^{7-p}\right]dx^{2}+\frac{1}{\left(\frac{u}{R}\right)^{\frac{7-p}{2}}\left[1-\left(\frac{u_{\Lambda}}{u}\right)^{7-p}\right]}du^{2}\end{equation}
 Following \cite{key-18} let us define a new coordinate\begin{equation}
\frac{u}{u_{\Lambda}}=1+\left(\frac{\rho}{R}\right)^{2}\end{equation}
 such that the singularity at $u=u_{\Lambda}$ maps to $\rho=0$.
Then for small $\rho$ the relevant $2d$ part of the Euclidean metric
($\tau=ix$) is given by\begin{equation}
ds^{2}\sim\left(7-p\right)u_{\Lambda}^{\frac{7-p}{2}}R^{-\frac{11-p}{2}}\rho^{2}d\tau^{2}+\frac{4}{7-p}u_{\Lambda}^{-\frac{3-p}{2}}R^{\frac{3-p}{2}}d\rho^{2}\end{equation}
 which can be written as \cite{key-19}\begin{equation}
ds^{2}\sim\Omega\left[d\rho^{2}+c^{2}\rho^{2}d\tau^{2}\right]\end{equation}
 where\begin{eqnarray}
\Omega & = & \frac{4}{7-p}u_{\Lambda}^{-\frac{3-p}{2}}R^{\frac{3-p}{2}}\\
c^{2} & = & \frac{\left(7-p\right)u_{\Lambda}^{\frac{7-p}{2}}R^{-\frac{11-p}{2}}}{\Omega}=\left(\frac{7-p}{2}\right)^{2}u_{\Lambda}^{5-p}R^{-\left(7-p\right)}\end{eqnarray}
 In order to avoid a conical singularity the coordinate $c\tau$ must
be identified we a periodicity of $2\pi$\begin{equation}
c\tau\rightarrow c\tau+2\pi\end{equation}
 means the periodicity of $\tau$ is given by\begin{equation}
\beta=\frac{2\pi}{c}=\frac{4\pi}{7-p}u_{\Lambda}^{-\frac{5-p}{2}}R^{\frac{7-p}{2}}\end{equation}

In a similar manner, the periodicity of the compactified coordinate
of the non critical $AdS_{n+1}$ background was calculated in \cite{key-7}
to be\begin{equation}
\beta=\frac{4\pi}{n}\frac{R^{2}}{u_{\Lambda}}\end{equation}

\section{Analytical expressions for energy and length of the string in near
extremal $D_{p}$ brane\label{sec:Analytical-expressions-for-Dp brane}}

In the case of near extremal $D_{p}$ brane, the metric is of the
form

\begin{equation}
\begin{aligned}f^{2}\left(u\right) & =\left(\frac{1}{2\pi\alpha'}\right)^{2}\left(\frac{u}{R_{AdS}}\right)^{7-p}\\
g^{2}\left(u\right) & =\left(\frac{1}{2\pi\alpha'}\right)^{2}\frac{1}{1-\left(\frac{u_{\Lambda}}{u}\right)^{7-p}}\end{aligned}
\end{equation}
 Inserting the above functions into the expression of the string's
length \eqref{eq:length} we have

\begin{equation}
\begin{aligned}\ell\left(u_{0}\right) & =2\frac{1}{f\left(u_{0}\right)}\int_{u_{0}}^{u_{b}}du\frac{f^{2}\left(u_{0}\right)}{f^{2}\left(u\right)}g\left(u\right)\frac{1}{\sqrt{1-\frac{f^{2}\left(u_{0}\right)}{f^{2}\left(u\right)}}}\\
 & =2\left(\frac{R_{AdS}}{u_{\Lambda}}\right)^{\frac{7-p}{2}}\int_{u_{0}}^{\infty}du\left(\frac{u_{0}}{u}\right)^{7-p}\frac{1}{\sqrt{1-\left(\frac{u_{\Lambda}}{u}\right)^{7-p}}\sqrt{1-\left(\frac{u_{0}}{u}\right)^{7-p}}}\end{aligned}
\end{equation}
 Define new variables\begin{eqnarray}
y & = & \frac{u}{u_{0}}\\
z & = & \left(\frac{u_{\Lambda}}{u_{0}}\right)^{7-p}\end{eqnarray}
 we can write the length as a function of $z$\begin{equation}
\ell\left(z\right)=2u_{\Lambda}\cdot\left(\frac{R_{AdS}}{u_{\Lambda}}\right)^{\frac{7-p}{2}}\cdot\left(z\right)^{\frac{5-p}{2\left(7-p\right)}}\cdot\int_{1}^{\infty}dy\frac{y^{-\left(7-p\right)}}{\sqrt{1-zy^{-\left(7-p\right)}}\sqrt{1-y^{-\left(7-p\right)}}}\end{equation}
 Notice that the length is measured in units of $u_{\Lambda}\cdot\left(\frac{R_{AdS}}{u_{\Lambda}}\right)^{\frac{7-p}{2}}$.
Then we express it in terms of the universal system of units defined
in \eqref{sec:Units-of-measurement} using the dictionary derived
in section \eqref{sec:Comparison-with-lattice} (table \eqref{Table:Dictionary of translation})
to get\begin{equation}
\ell\left(z\right)=\frac{2d_{p}}{m_{gb}}\left[\left(z\right)^{\frac{5-p}{2\left(7-p\right)}}\int_{1}^{\infty}dy\frac{y^{-\left(7-p\right)}}{\sqrt{1-zy^{-\left(7-p\right)}}\sqrt{1-y^{-\left(7-p\right)}}}\right]\end{equation}
 where $d_{p}=3.4,\,2.24$ for $p=3,4$, respectively. In a similar
way we calculate the $\mathcal{K}$ function \eqref{eq:Kappa}\begin{equation}
\begin{aligned}\mathcal{K}\left(u_{0}\right) & =\int_{u_{0}}^{u_{b}}du\, g\left(u\right)\left(1-\sqrt{1-\frac{f^{2}\left(u_{0}\right)}{f^{2}\left(u\right)}}\right)+\int_{u_{\Lambda}}^{u_{0}}du\, g\left(u\right)\\
 & =\frac{1}{2\pi\alpha'}\left[\int_{u_{0}}^{\infty}du\,\frac{1}{\sqrt{1-\left(\frac{u_{\Lambda}}{u}\right)^{7-p}}}\left(1-\sqrt{1-\left(\frac{u_{0}}{u}\right)^{7-p}}\right)+\int_{u_{\Lambda}}^{u_{0}}du\,\frac{1}{\sqrt{1-\left(\frac{u_{\Lambda}}{u}\right)^{7-p}}}\right]\end{aligned}
\end{equation}

\begin{equation}
\mathcal{K}\left(z\right)=\frac{u_{\Lambda}}{\alpha'}z^{-\frac{1}{7-p}}\frac{1}{2\pi}\left[\int_{1}^{\infty}dy\,\frac{1-\sqrt{1-y^{-\left(7-p\right)}}}{\sqrt{1-zy^{-\left(7-p\right)}}}+\int_{z^{\frac{1}{7-p}}}^{1}dy\,\frac{1}{\sqrt{1-zy^{-\left(7-p\right)}}}\right]\end{equation}
 Notice that the energy, which has the same dimension as $\mathcal{K}$,
is measured in units of $\frac{u_{\Lambda}}{\alpha'}$. Expressing
it in terms of the universal system of units we have\begin{equation}
\mathcal{K}\left(z\right)=\frac{\sigma}{m_{gb}}d_{p}z^{-\frac{1}{7-p}}\left[\int_{1}^{\infty}dy\,\frac{1-\sqrt{1-y^{-\left(7-p\right)}}}{\sqrt{1-zy^{-\left(7-p\right)}}}+\int_{z^{\frac{1}{7-p}}}^{1}dy\,\frac{1}{\sqrt{1-zy^{-\left(7-p\right)}}}\right]\end{equation}
 Finally, the energy is given by \eqref{eq:Energy}

\begin{equation}
E\left(u_{0}\right)=f\left(u_{0}\right)\cdot\ell\left(u_{0}\right)-2\mathcal{K}\left(u_{0}\right)\end{equation}
 \begin{equation}
\begin{aligned}E\left(z\right) & =\frac{1}{2\pi\alpha'}\left(\frac{u_{\Lambda}}{R_{AdS}}\right)^{\frac{7-p}{2}}z^{-\frac{1}{2}}\cdot\ell\left(z\right)-2\mathcal{K}\left(z\right)\\
 & =\sigma z^{-\frac{1}{2}}\cdot\ell\left(z\right)-2\mathcal{K}\left(z\right)\\
 & =\frac{\sigma}{m_{gb}}2d_{p}z^{-\frac{1}{7-p}}\left[\int_{1}^{\infty}dy\left[\frac{1-\sqrt{1-y^{-\left(7-p\right)}}}{\sqrt{1-y^{-\left(7-p\right)}}\sqrt{1-zy^{-\left(7-p\right)}}}\right]-\int_{z^{\frac{1}{7-p}}}^{1}dy\,\frac{1}{\sqrt{1-zy^{-\left(7-p\right)}}}\right]\end{aligned}
\end{equation}

\section{The values of the constant terms in near extremal $D_{p}$ brane\label{sec:The-values-of-constants-DpBranes}}

In this appendix we will find the constant terms of the potential
at short and long distances for the near extremal $D_{p}$ brane,
using the expressions for the energy and length derived in appendix
\eqref{sec:Analytical-expressions-for-Dp brane} using the mass subtraction
regularization scheme. Pay attention that we are before fixing the
zero energy point as in \cite{key-9}'s notations. First, at short
distances, i.e. the limit $z\rightarrow0$, one finds\begin{equation}
E\rightarrow\frac{\sigma}{m_{gb}}2d_{p}\left[z^{-\frac{1}{7-p}}\int_{1}^{\infty}dy\left(\frac{1}{\sqrt{1-y^{-\left(7-p\right)}}}-1\right)-\frac{\sqrt{\pi}\Gamma\left[1-\frac{1}{7-p}\right]}{\Gamma\left[\frac{1}{2}-\frac{1}{7-p}\right]}+\ldots\right]\end{equation}
 where the three dots denote terms which vanish in this limit. The
length is given by\begin{equation}
\ell\rightarrow\frac{1}{m_{gb}}\left(z\right)^{\frac{5-p}{2\left(7-p\right)}}\left[2d_{p}\int_{1}^{\infty}dy\frac{y^{-\left(7-p\right)}}{\sqrt{1-y^{-\left(7-p\right)}}}\right]\end{equation}
 such that\begin{equation}
E=\left[\#_{p}\right]\cdot\sigma\left(m_{gb}\right)^{-\frac{7-p}{5-p}}\cdot\ell^{-\frac{2}{5-p}}-\frac{\sigma}{m_{gb}}\frac{2d_{p}\sqrt{\pi}\Gamma\left[1-\frac{1}{7-p}\right]}{\Gamma\left[\frac{1}{2}-\frac{1}{7-p}\right]}\end{equation}
 In fact, we could alreay conclude this power law behavior by the
analysis in \eqref{sec:asymptotically coulombian appendix}, but the
important point here is the value of the constant term\begin{equation}
C_{S}=-\frac{\sigma}{m_{gb}}\frac{2d_{p}\sqrt{\pi}\Gamma\left[1-\frac{1}{7-p}\right]}{\Gamma\left[\frac{1}{2}-\frac{1}{7-p}\right]}\end{equation}
 Next we examine the behavior at long distances, i.e. $z\rightarrow1$,
there\begin{equation}
E=\sigma\cdot\ell-2\mathcal{K}\left(z=1\right)\end{equation}
 such that the constant term is given by\begin{equation}
\begin{aligned}C_{L} & =-2\mathcal{K}\left(z=1\right)\\
 & =-\frac{\sigma}{m_{gb}}2d_{p}\left[\int_{1}^{\infty}dy\,\left(\frac{1}{\sqrt{1-y^{-\left(7-p\right)}}}-1\right)\right]\end{aligned}
\end{equation}
 Finally, we conclude that the relative constant will be equal to\begin{equation}
\begin{aligned}\Delta C & =C_{L}-C_{S}\\
 & =-\frac{\sigma}{m_{gb}}2d_{p}\left[\frac{\sqrt{\pi}\Gamma\left[1-\frac{1}{7-p}\right]}{\Gamma\left[\frac{1}{2}-\frac{1}{7-p}\right]}+\int_{1}^{\infty}dy\,\left(\frac{1}{\sqrt{1-y^{-\left(7-p\right)}}}-1\right)\right]\end{aligned}
\end{equation}
 The expression in the square brackets appears to be equal to one
such that\[
\Delta C=-2d_{p}\frac{\sigma}{m_{gb}}\]

\begin{center}
\begin{tabular}{|c|c|c|}
\hline
$p$  & 3  & 4 \tabularnewline
\hline
\hline
$\Delta C\left[\frac{\sigma}{m_{gb}}\right]$  & -6.8  & -4.48\tabularnewline
\hline
\end{tabular}
\par\end{center}

\section{Analytical expressions for energy and length of the string in non-critical
$AdS_{n+1}$\label{sec:Analytical-expressions-for-AdS}}

Following the same steps as for the $D$ branes example \eqref{sec:Analytical-expressions-for-Dp brane},
the length of the string in the non-critical $AdS_{n+1}$ case is
given by \begin{equation}
\ell\left(u_{0}\right)=2\frac{R_{AdS}^{2}}{u_{0}^{2}}\int_{u_{0}}^{\infty}du\frac{1}{\left(\frac{u}{u_{0}}\right)^{4}\sqrt{1-\left(\frac{u_{\Lambda}}{u}\right)^{n}}\sqrt{1-\left(\frac{u_{0}}{u}\right)^{4}}}\end{equation}
 Changing variables to $y\equiv\frac{u}{u_{0}}$, $z\equiv\left(\frac{u_{\Lambda}}{u_{0}}\right)^{n}$
we have \begin{equation}
=2\frac{R_{AdS}^{2}}{u_{\Lambda}}z^{\frac{1}{n}}\int_{1}^{\infty}dy\frac{y^{-4}}{\sqrt{1-zy^{-n}}\sqrt{1-y^{-4}}}\end{equation}
 Notice that in the present case length is measured in units of $\frac{R_{AdS}^{2}}{u_{\Lambda}}$.
Then expressing it in terms of the universal system of units defined
in \eqref{sec:Units-of-measurement} using the dictionary derived
in section \eqref{sec:Comparison-with-lattice} (table \eqref{Table:Dictionary of translation})
we result with\begin{equation}
\ell\left(z\right)=\frac{1}{m_{gb}}\left[5.04z^{\frac{1}{n}}\int_{1}^{\infty}dy\frac{y^{-4}}{\sqrt{1-zy^{-n}}\sqrt{1-y^{-4}}}\right]\end{equation}
 In the same manner we derive the explicit expression for the $\mathcal{K}$
function

\begin{equation}
\begin{aligned}\mathcal{K}\left(u_{0}\right) & =\frac{1}{2\pi\alpha'}\int_{u_{0}}^{\infty}du\frac{1}{\sqrt{1-\left(\frac{u_{\Lambda}}{u}\right)^{n}}}\left(1-\sqrt{1-\left(\frac{u_{0}}{u}\right)^{4}}\right)+\frac{1}{2\pi}\int_{u_{\Lambda}}^{u_{0}}du\frac{1}{\sqrt{1-\left(\frac{u_{\Lambda}}{u}\right)^{n}}}\\
 & =\frac{u_{\Lambda}}{\alpha'}z^{-\frac{1}{n}}\frac{1}{2\pi}\left[\int_{1}^{\infty}dy\frac{1}{\sqrt{1-zy^{-n}}}\left(1-\sqrt{1-y^{-4}}\right)+\frac{1}{2\pi}\int_{z^{\frac{1}{n}}}^{1}dy\frac{1}{\sqrt{1-zy^{-n}}}\right]\end{aligned}
\end{equation}
 We see that as in the $D$ branes case, energy is measured in units
of $\frac{u_{\Lambda}}{\alpha'}$. In the universal system of units
the $\mathcal{\mathcal{K}}$ function will take the form\begin{equation}
\mathcal{K}\left(z\right)=\frac{\sigma}{m_{gb}}\left[2.52z^{-\frac{1}{n}}\left(\int_{1}^{\infty}dy\frac{1}{\sqrt{1-zy^{-n}}}\left(1-\sqrt{1-y^{-4}}\right)+\int_{z^{\frac{1}{n}}}^{1}dy\frac{1}{\sqrt{1-zy^{-n}}}\right)\right]\end{equation}
 Finally the expression for the energy is given by

\begin{equation}
E\left(z\right)=\sigma z^{-\frac{2}{n}}\cdot\ell\left(z\right)-2\mathcal{K}\left(z\right)\end{equation}
 such that\begin{equation}
E\left(z\right)=\frac{\sigma}{m_{gb}}5.04z^{-\frac{1}{n}}\left[\int_{1}^{\infty}dy\frac{1-\sqrt{1-y^{-4}}}{\sqrt{1-zy^{-n}}\sqrt{1-y^{-4}}}-\int_{z^{\frac{1}{n}}}^{1}dy\frac{1}{\sqrt{1-zy^{-n}}}\right]\end{equation}

\section{The values of the constant terms in non-critical $AdS_{n+1}$\label{sec:The-value-of-the-constant-AdS}}

Following again the same steps as for the $D$ brane cases in \eqref{sec:The-values-of-constants-DpBranes},
we will find the constant terms of the energy in the cases of non-critical
$AdS_{n+1}$. We will use the expressions for the energy and length
derived in appendix \eqref{sec:Analytical-expressions-for-AdS}. Pay
attention that we are before fixing the zero energy point as in \cite{key-9}'s
notations. First, at short distances, i.e. the limit $z\rightarrow0$,
one finds

\begin{equation}
E\left(z\right)\rightarrow\frac{\sigma}{m_{gb}}5.04\left[z^{-\frac{1}{n}}\int_{1}^{\infty}dy\left(\frac{1}{\sqrt{1-y^{-4}}}-1\right)-\frac{\sqrt{\pi}\Gamma\left[1-\frac{1}{n}\right]}{\Gamma\left[\frac{1}{2}-\frac{1}{n}\right]}+\ldots\right]\end{equation}
 where the three dots denote terms which vanish in this limit. The
length is given by\begin{equation}
\ell\left(z\right)\rightarrow\frac{1}{m_{gb}}5.04z^{\frac{1}{n}}\left[\int_{1}^{\infty}dy\frac{y^{-4}}{\sqrt{1-y^{-4}}}\right]\end{equation}
 such that\begin{equation}
E=\frac{\sigma}{m_{gb}^{2}}\left(5.04\right)^{2}\left[\int_{1}^{\infty}dy\frac{y^{-4}}{\sqrt{1-y^{-4}}}\right]\left[\int_{1}^{\infty}dy\left(\frac{1}{\sqrt{1-y^{-4}}}-1\right)-1\right]\frac{1}{\ell}-\frac{\sigma}{m_{gb}}\frac{5.04\sqrt{\pi}\Gamma\left[1-\frac{1}{n}\right]}{\Gamma\left[\frac{1}{2}-\frac{1}{n}\right]}\end{equation}
 \begin{equation}
E\cong-\left(0.36\cdot\left(5.04\right)^{2}\frac{\sigma}{m_{gb}^{2}}\right)\frac{1}{\ell}-\frac{\sigma}{m_{gb}}\frac{5.04\sqrt{\pi}\Gamma\left[1-\frac{1}{n}\right]}{\Gamma\left[\frac{1}{2}-\frac{1}{n}\right]}\end{equation}
 We find that the constant term at short distances is given by\begin{equation}
C_{S}=-\frac{\sigma}{m_{gb}}\frac{5.04\sqrt{\pi}\Gamma\left[1-\frac{1}{n}\right]}{\Gamma\left[\frac{1}{2}-\frac{1}{n}\right]}\end{equation}
 In the opposite limit, $z\rightarrow1$, corresponding to large distances,
the energy takes the form\begin{equation}
E=\sigma\cdot\ell-2\mathcal{K}\left(z=1\right)\end{equation}
 such that the constant term is equal to\begin{equation}
\begin{aligned}C_{L} & =-2\mathcal{K}\left(z=1\right)\\
 & =-\frac{\sigma}{m_{gb}}\left[5.04\left(\int_{1}^{\infty}dy\frac{1-\sqrt{1-y^{-4}}}{\sqrt{1-y^{-n}}}\right)\right]\end{aligned}
\end{equation}
 Finally, we conclude that the relative constant will be equal to\begin{equation}
\begin{aligned}\Delta C & =C_{L}-C_{S}\\
 & =-\frac{\sigma}{m_{gb}}5.04\left[\frac{\sqrt{\pi}\Gamma\left[1-\frac{1}{n}\right]}{\Gamma\left[\frac{1}{2}-\frac{1}{n}\right]}+\int_{1}^{\infty}dy\frac{1-\sqrt{1-y^{-4}}}{\sqrt{1-y^{-n}}}\right]\end{aligned}
\end{equation}
 and this can be evaluate numerically for different values of $n$:

\begin{center}
\[
\Delta C=-5.34\]

\par\end{center}

\section{The Wilson line of the Klebanov Strassler background\label{sec:The-potential-for-KS}}

The supergravity solution of the deformed conifold is of the following
form \cite{key-13,Benna:2007mb} \begin{equation}
ds^{2}=h^{-\frac{1}{2}}\left(\tau\right)dx_{0123}^{2}+h^{\frac{1}{2}}\left(\tau\right)ds_{6}^{2}\end{equation}
 where $ds_{6}^{2}$ is the metric of the deformed conifold\begin{equation}
ds_{6}^{2}=\frac{\epsilon^{4/3}}{2}K\left(\tau\right)\left[\frac{1}{3K^{3}\left(\tau\right)}\left(d\tau^{2}+\left(g^{5}\right)^{2}\right)+\cosh^{2}\left(\frac{\tau}{2}\right)\left[\left(g^{3}\right)^{2}+\left(g^{4}\right)^{2}\right]+\sinh^{2}\left(\frac{\tau}{2}\right)\left[\left(g^{1}\right)^{2}+\left(g^{2}\right)^{2}\right]\right]\end{equation}
 $\epsilon$ is the energy scale and the functions $h\left(\tau\right)$
and $K\left(\tau\right)$ are given by\begin{equation}
h\left(\tau\right)=\left(g_{s}M\alpha'\right)^{2}2^{2/3}\epsilon^{-8/3}I\left(\tau\right)\end{equation}
 \begin{equation}
I\left(\tau\right)\equiv\int_{\tau}^{\infty}dx\frac{x\coth x-1}{\sinh^{2}x}\left(\sinh2x-2x\right)^{\frac{1}{3}}\end{equation}
 \begin{equation}
K\left(\tau\right)=\frac{\left(\sinh\left(2\tau\right)-2\tau\right)^{\frac{1}{3}}}{2^{\frac{1}{3}}\sinh\tau}\end{equation}
 Therefore the $f$ and $g$ functions are\begin{equation}
\begin{aligned}f\left(\tau\right) & =h^{-\frac{1}{2}}\left(\tau\right)\\
g\left(\tau\right) & =\frac{1}{\sqrt{6}}\epsilon^{2/3}K^{-1}\left(\tau\right)\end{aligned}
\end{equation}
 Near the boundary, i.e. at large $\tau$, the $h\left(\tau\right)$
and $K\left(\tau\right)$ function takes the form\begin{equation}
h\left(\tau\gg1\right)=\left(g_{s}M\alpha'\right)^{2}3\cdot2^{2/3}\epsilon^{-8/3}\tau e^{-\frac{4}{3}\tau}\end{equation}
 \begin{equation}
K\left(\tau\gg1\right)=2^{\frac{1}{3}}e^{-\frac{1}{3}\tau}\end{equation}
 Then the distance between the quarks is\begin{equation}
\begin{aligned}\ell\left(\tau_{0}\right) & =\frac{2}{f\left(\tau_{0}\right)}\int_{\tau_{0}}^{\infty}d\tau\, g\left(\tau\right)\left[\frac{f\left(\tau_{0}\right)}{f\left(\tau\right)}\right]^{2}\frac{1}{\sqrt{1-\left[\frac{f\left(\tau_{0}\right)}{f\left(\tau\right)}\right]^{2}}}\\
 & =\frac{2^{\frac{2}{3}}\epsilon^{2/3}}{\sqrt{6}f\left(\tau_{0}\right)}\int_{\tau_{0}}^{\infty}d\tau\, e^{\frac{1}{3}\tau}\left(\frac{\tau}{\tau_{0}}e^{-\frac{4}{3}\left(\tau-\tau_{0}\right)}\right)\frac{1}{\sqrt{1-\frac{\tau}{\tau_{0}}e^{-\frac{4}{3}\left(\tau-\tau_{0}\right)}}}\\
\left(t=\frac{\tau}{\tau_{0}}\right) & =\frac{2^{\frac{2}{3}}\epsilon^{2/3}}{\sqrt{6}f\left(\tau_{0}\right)}e^{+\frac{1}{3}\tau_{0}}\tau_{0}\int_{1}^{\infty}dt\,\left(te^{-\tau_{0}\left(t-1\right)}\right)\frac{1}{\sqrt{1-te^{-\frac{4}{3}\tau_{0}\left(t-1\right)}}}\end{aligned}
\end{equation}
 In the limit $\tau_{0}\rightarrow\infty$ the integral takes the
value\begin{equation}
\tau_{0}\int_{1}^{\infty}dt\left(te^{-\tau_{0}\left(t-1\right)}\right)\frac{1}{\sqrt{1-te^{-\frac{4}{3}\tau_{0}\left(t-1\right)}}}\rightarrow1.8\end{equation}
 such that\begin{equation}
\ell\left(\tau_{0}\right)=\frac{2^{\frac{2}{3}}\cdot1.8\cdot\epsilon^{2/3}}{\sqrt{6}f\left(\tau_{0}\right)}e^{+\frac{1}{3}\tau_{0}}=2^{1/2}\cdot1.8\cdot\left(g_{s}M\alpha'\right)\epsilon^{-2/3}\tau_{0}^{1/2}e^{-\frac{1}{3}\tau_{0}}\end{equation}
 The energy of the pair is\begin{equation}
E\left(\tau_{0}\right)=f\left(\tau_{0}\right)\ell\left(\tau_{0}\right)-2\mathcal{K}\left(\tau_{0}\right)\end{equation}
 where the $\mathcal{K}$ function is\begin{equation}
\begin{aligned}\mathcal{K}\left(\tau_{0}\right) & =\int_{\tau_{0}}^{\infty}d\tau\, g\left(\tau\right)\left[1-\sqrt{1-\left[\frac{f\left(\tau_{0}\right)}{f\left(\tau\right)}\right]^{2}}\right]+\int_{0}^{\tau_{0}}d\tau\, g\left(\tau\right)\\
 & =\frac{\epsilon^{2/3}}{2^{\frac{1}{3}}\sqrt{6}}\int_{\tau_{0}}^{\infty}d\tau\, e^{\frac{1}{3}\tau}\left[1-\sqrt{1-\frac{\tau}{\tau_{0}}e^{-\frac{4}{3}\left(\tau-\tau_{0}\right)}}\right]+\frac{\epsilon^{2/3}}{2^{\frac{1}{3}}\sqrt{6}}\int_{0}^{\tau_{0}}d\tau\, e^{\frac{1}{3}\tau}\\
\left(t=\frac{\tau}{\tau_{0}}\right) & =\frac{\epsilon^{2/3}}{2^{\frac{1}{3}}\sqrt{6}}\tau_{0}\int_{1}^{\infty}dt\, e^{\frac{1}{3}\tau_{0}t}\left[1-\sqrt{1-te^{-\frac{4}{3}\tau_{0}\left(t-1\right)}}\right]+\frac{3\epsilon^{2/3}}{2^{\frac{1}{3}}\sqrt{6}}\left(e^{\frac{1}{3}\tau_{0}}-1\right)\end{aligned}
\end{equation}
 Let us write this expression in the next form\begin{equation}
=\frac{\epsilon^{2/3}}{2^{\frac{1}{3}}\sqrt{6}}e^{\frac{1}{3}\tau_{0}}\tau_{0}\int_{1}^{\infty}dt\, e^{\frac{1}{3}\tau_{0}\left(t-1\right)}\left[1-\sqrt{1-te^{-\frac{4}{3}\tau_{0}\left(t-1\right)}}\right]+\frac{3\epsilon^{2/3}}{2^{\frac{1}{3}}\sqrt{6}}\left(e^{\frac{1}{3}\tau_{0}}-1\right)\end{equation}
 In the limit $\tau_{0}\rightarrow\infty$ the integral takes the
value\begin{equation}
\tau_{0}\int_{1}^{\infty}dt\, e^{\frac{1}{3}\tau_{0}\left(t-1\right)}\left[1-\sqrt{1-te^{-\frac{4}{3}\tau_{0}\left(t-1\right)}}\right]\rightarrow0.6\end{equation}
 such that the Kappa function will take the form\begin{equation}
\mathcal{K}\left(\tau_{0}\right)=\frac{3.6\epsilon^{2/3}}{2^{\frac{1}{3}}\sqrt{6}}e^{\frac{1}{3}\tau_{0}}-\frac{3\epsilon^{2/3}}{2^{\frac{1}{3}}\sqrt{6}}\end{equation}
 Thus the energy in the region close to the boundary is given by\begin{equation}
\frac{E}{\epsilon^{2/3}}=-\frac{1.8\cdot2^{\frac{2}{3}}}{\sqrt{6}}e^{+\frac{1}{3}\tau_{0}}+\frac{3\cdot2^{\frac{2}{3}}}{\sqrt{6}}\end{equation}

We would like to express the energy as a function of the distance
between the quarks\begin{equation}
\ln\left[\left(g_{s}M\alpha'\right)^{-1}\epsilon^{2/3}\ell\right]=-\frac{1}{3}\tau_{0}+\ln\left(\tau_{0}^{\frac{1}{2}}\right)+\ln\left(2^{1/2}\cdot1.8\right)\label{eq:KS.Length.exact}\end{equation}
 To first order (at large $\tau_{0}$) we have\begin{equation}
\ln\left[\left(g_{s}M\alpha'\right)^{-1}\epsilon^{2/3}\ell\right]\cong-\frac{1}{3}\tau_{0}\label{eq:KS.leadingLengthBehavior}\end{equation}
 such that the leading behavior of the energy is indeed Coulombic\begin{equation}
\frac{E}{\epsilon^{2/3}}\cong-\frac{1.8\cdot2^{\frac{2}{3}}}{\sqrt{6}}\frac{\left(g_{s}M\alpha'\right)}{\epsilon^{2/3}\ell}+\frac{3\cdot2^{\frac{2}{3}}}{\sqrt{6}}\end{equation}
 It is not a surprise since what we have done by taking the leading
behavior \eqref{eq:KS.leadingLengthBehavior} is exactly considering
only the leading geometry, which is $AdS$ , as discussed in section
\eqref{sec:the wilson line of the holographic models}.

We now want to estimate the leading correction to the Coulombic potential.
To do that we first find the leading correction to the dependence
of the length on $\tau_{0}$ by plugging in the leading order \eqref{eq:KS.leadingLengthBehavior}
into the subleaing term of \eqref{eq:KS.Length.exact}\begin{equation}
\ln\left[\left(g_{s}M\alpha'\right)^{-1}\epsilon^{2/3}\ell\right]\cong-\frac{1}{3}\tau_{0}+\ln\left(\sqrt{-\ln\left[\left(g_{s}M\alpha'\right)^{-1}\epsilon^{2/3}\ell\right]}\right)+\ln\left(\sqrt{3}\cdot2^{-\frac{2}{3}}\cdot1.8\right)\end{equation}
 \begin{equation}
\frac{1}{3}\tau_{0}\cong-\ln\left[\left(g_{s}M\alpha'\right)^{-1}\epsilon^{2/3}\ell\right]+\ln\left(\sqrt{-\ln\left[\left(g_{s}M\alpha'\right)^{-1}\epsilon^{2/3}\ell\right]}\right)+\ln\left(\sqrt{3}\cdot2^{-\frac{2}{3}}\cdot1.8\right)\end{equation}
 Therefore the energy will take the form\begin{equation}
\frac{E}{\epsilon^{2/3}}\cong-\frac{1.8^{2}}{2}\frac{\sqrt{-\ln\left[\left(g_{s}M\alpha'\right)^{-1}\epsilon^{2/3}\ell\right]}}{\left(g_{s}M\alpha'\right)^{-1}\epsilon^{2/3}\ell}+\frac{3\cdot2^{\frac{2}{3}}}{\sqrt{6}}\label{eq:KS.Potential.Small}\end{equation}

Until now we have worked in the natural system of units of the KS
model, but in order to make a comparison we have to switch to the
universal system, as follows from the dictionary derived in section
\eqref{sec:Comparison-with-lattice} (table \eqref{Table:Dictionary of translation})
\begin{equation}
\begin{aligned}f\left(\tau\right) & =\left[2^{-\frac{1}{3}}I^{-\frac{1}{2}}\left(\tau\right)\right]\frac{\epsilon^{\frac{4}{3}}}{g_{s}M\alpha'}\\
 & =\left[\frac{1.51}{0.86}2^{-\frac{1}{3}}I^{-\frac{1}{2}}\left(\tau\right)\right]\sigma\\
g\left(\tau\right) & =\left[\frac{1}{\sqrt{6}}K^{-1}\left(\tau\right)\right]\epsilon^{2/3}\\
 & =\left[\frac{1.51}{\sqrt{6}}K^{-1}\left(\tau\right)\right]\frac{\sigma}{m_{gb}}\end{aligned}
\end{equation}

The constant term of the potential in the small distances regime can
be read off from \eqref{eq:KS.Potential.Small}

\begin{equation}
C_{S}=\frac{3\cdot2^{\frac{2}{3}}}{\sqrt{6}}\epsilon^{2/3}\cong1.94\epsilon^{2/3}=2.93\frac{\sigma}{m_{gb}}\end{equation}
 On the other hand, the constant term at large distances can be calculated
numerically to be\begin{equation}
C_{L}=-2\mathcal{K}\left(\tau=0\right)=-1.44\frac{\sigma}{m_{gb}}\end{equation}
 Therefore the relative constant is given by

\begin{equation}
\Delta C=C_{L}-C_{S}=-4.37\frac{\sigma}{m_{gb}}\end{equation}

%\cite{Athenodorou:2007du}

\end{document}